\documentclass[conference,10pt]{IEEEtran}
\IEEEoverridecommandlockouts

\usepackage{preamble}
\usepackage{definitions}

\def\BibTeX{{\rm B\kern-.05em{\sc i\kern-.025em b}\kern-.08em
    T\kern-.1667em\lower.7ex\hbox{E}\kern-.125emX}}
\begin{document}

\title{
\sysname: Collision Detector Bug Discovery in Autonomous Driving Simulators
}

\author{
\IEEEauthorblockN{Weiwei Fu}
\IEEEauthorblockA{\textit{City University of Hong Kong} \\
Hong Kong, China \\
weiweifu2-c@my.cityu.edu.hk}
\and
\IEEEauthorblockN{Heqing Huang}
\IEEEauthorblockA{\textit{City University of Hong Kong} \\
Hong Kong, China \\
heqhuang@cityu.edu.hk}
\and
\IEEEauthorblockN{Yifan Zhang}
\IEEEauthorblockA{\textit{ETH Zurich} \\
Zurich, Switzerland \\
yifan.zhang@ivt.baug.ethz.ch}
\and
\IEEEauthorblockN{Ke Zhang}
\IEEEauthorblockA{\textit{Chinese University of Hong Kong} \\
Hong Kong, China \\
zk019@ie.cuhk.edu.hk}
\and
\IEEEauthorblockN{Jin Huang}
\IEEEauthorblockA{\textit{Tsinghua University} \\
Beijing, China \\
huangjin@tsinghua.edu.cn}
\and
\IEEEauthorblockN{Wei-Bin Lee}
\IEEEauthorblockA{\textit{Hon Hai Research Institute} \\
Taipei, China \\
wei-bin.lee@foxconn.com}
 \and
\IEEEauthorblockN{Jianping Wang}
\IEEEauthorblockA{\textit{City University of Hong Kong} \\
Hong Kong, China \\
jianwang@cityu.edu.hk}

}

\pagestyle{plain}

\maketitle

\begin{abstract}


With the increasing adoption of autonomous vehicles, ensuring the reliability of autonomous driving systems (ADSs) deployed on autonomous vehicles has become a significant concern. Driving simulators have emerged as crucial platforms for testing autonomous driving systems, offering realistic, dynamic, and configurable environments. However, existing simulation-based ADS testers have largely overlooked the reliability of the simulators, potentially leading to overlooked violation scenarios and subsequent safety security risks during real-world deployment. In our investigations, we identified that collision detectors in simulators could fail to detect and report collisions in certain collision scenarios, referred to as \emph{ignored collision scenarios}.

This paper aims to systematically discover ignored collision scenarios to improve the reliability of autonomous driving simulators. To this end, we present ICSFuzz, a black-box fuzzing approach to discover ignored collision scenarios efficiently. 
Drawing upon the fact that the ignored collision scenarios are a sub-type of collision scenarios, our approach starts with the determined collision scenarios.
Following the guidance provided by empirically studied factors contributing to collisions, we selectively mutate arbitrary collision scenarios in a step-wise manner toward the ignored collision scenarios and effectively discover them.

We compare \sysname with DriveFuzz, a state-of-the-art simulation-based autonomous driving system testing method, by replacing its oracle with our ignored-collision-aware oracle.
The evaluation demonstrates that \sysname outperforms DriveFuzz by finding 10\textasciitilde20x more ignored collision scenarios with a 20\textasciitilde70x speedup.
Within the discovered ignored collision scenarios, there are seven more types of ignored collision scenarios that DriveFuzz did not find.
All the discovered ignored collisions have been confirmed by developers with one CVE ID assigned.

\end{abstract}


\vspace{-1mm}
\section{Introduction}
\label{sec:introduction}
The cornerstone of an autonomous vehicle is the reliability
of its ADS, e.g., Apollo~\cite{apollo} and Autoware~\cite{autoware}. 
Among all measures to test, validate, and improve the reliability of an ADS, 
driving simulators, e.g.,  Carla~\cite{carla} and LGSVL~\cite{lgsvl},  play an essential role by
providing a virtual environment where ADSs can be tested and evaluated extensively before real-world deployment. Simulators allow testers to simulate a wide range of scenarios and conditions, including complex and potentially
dangerous situations that may be difficult to replicate in real-world testing.  
Some testers~\cite{ABLE,zhou_specification,tang_systematic,avfuzzer,mosat,song2023discovering}
have successfully identified scenarios causing collision via the feedback of collision detectors in simulators.
For example, DriveFuzz~\cite{drivefuzz} uses Carla to detect rear-end collisions due to the deficiency of the motion planning module in Autoware.

\begin{figure}[tp]
\centering
\includegraphics[width=0.9\linewidth]{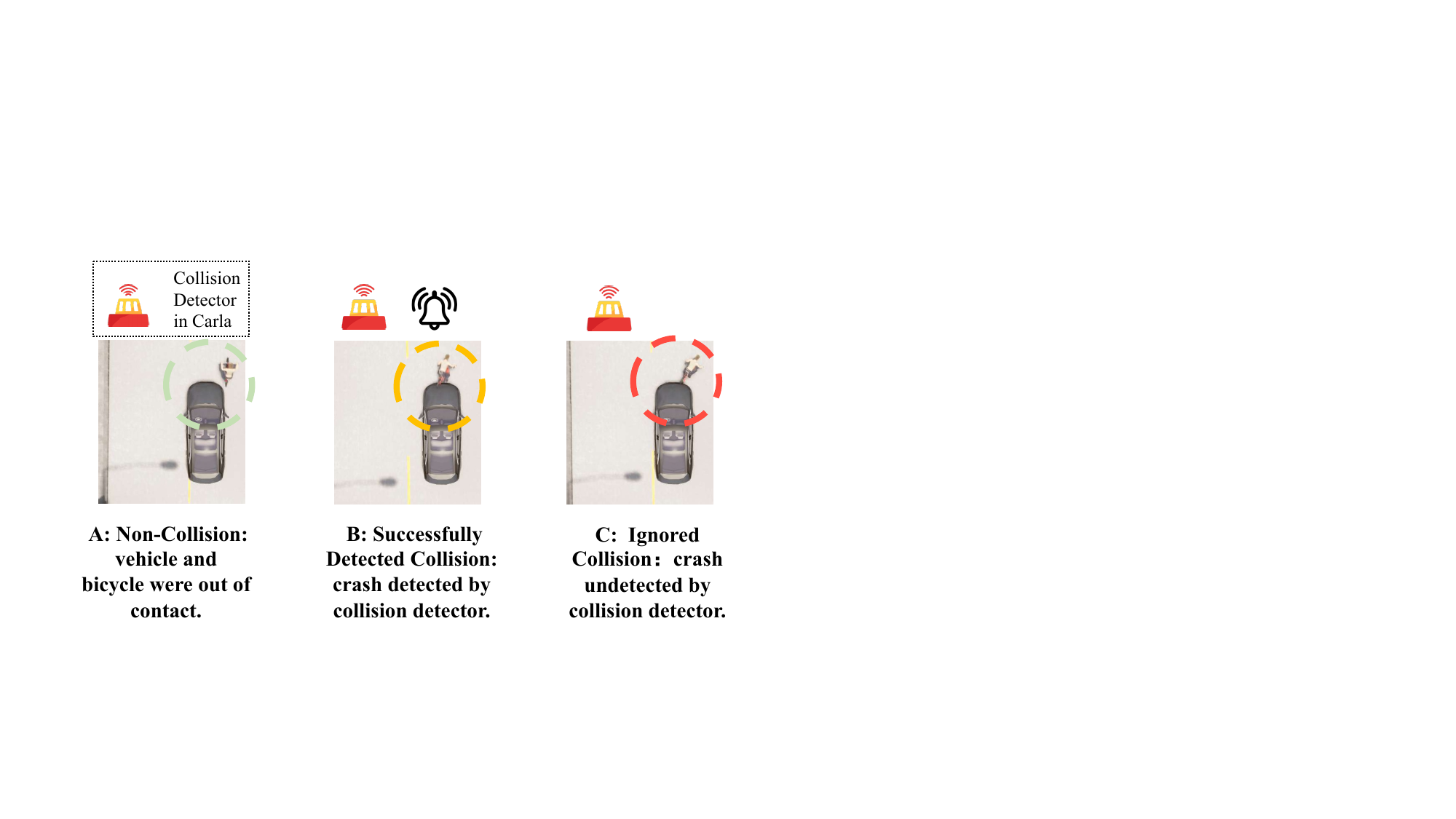} 
\caption{
Collision scenarios in  Carla, a driving simulator.
}
\label{fig:motivating example}
\vspace{-8mm}
\end{figure}

One implicit but fundamental assumption made in the existing simulation-driven testers is that the collision detector in a simulator is 100\% accurate where it always reports collision whenever a collision happens; otherwise, it reports non-collision, as illustrated in  Figure~\ref{fig:motivating example} (A) and (B). However, our experiments have shown that the collision detector in Carla may fail to detect some collisions, as shown in Figure~\ref{fig:motivating example} (C). The ignorance of such scenarios may lead to severe accidents in the real world. For example,  several Baidu Apollo Go autonomous vehicles crashed due to false negative reports from the simulators~\cite{Apollo_go_hit_and_run}. In this paper, we refer to the occurrences of false negatives in a collision detector as \emph{ignored collisions}.

\begin{figure}[t]
\centering
\includegraphics[width=0.9\linewidth]{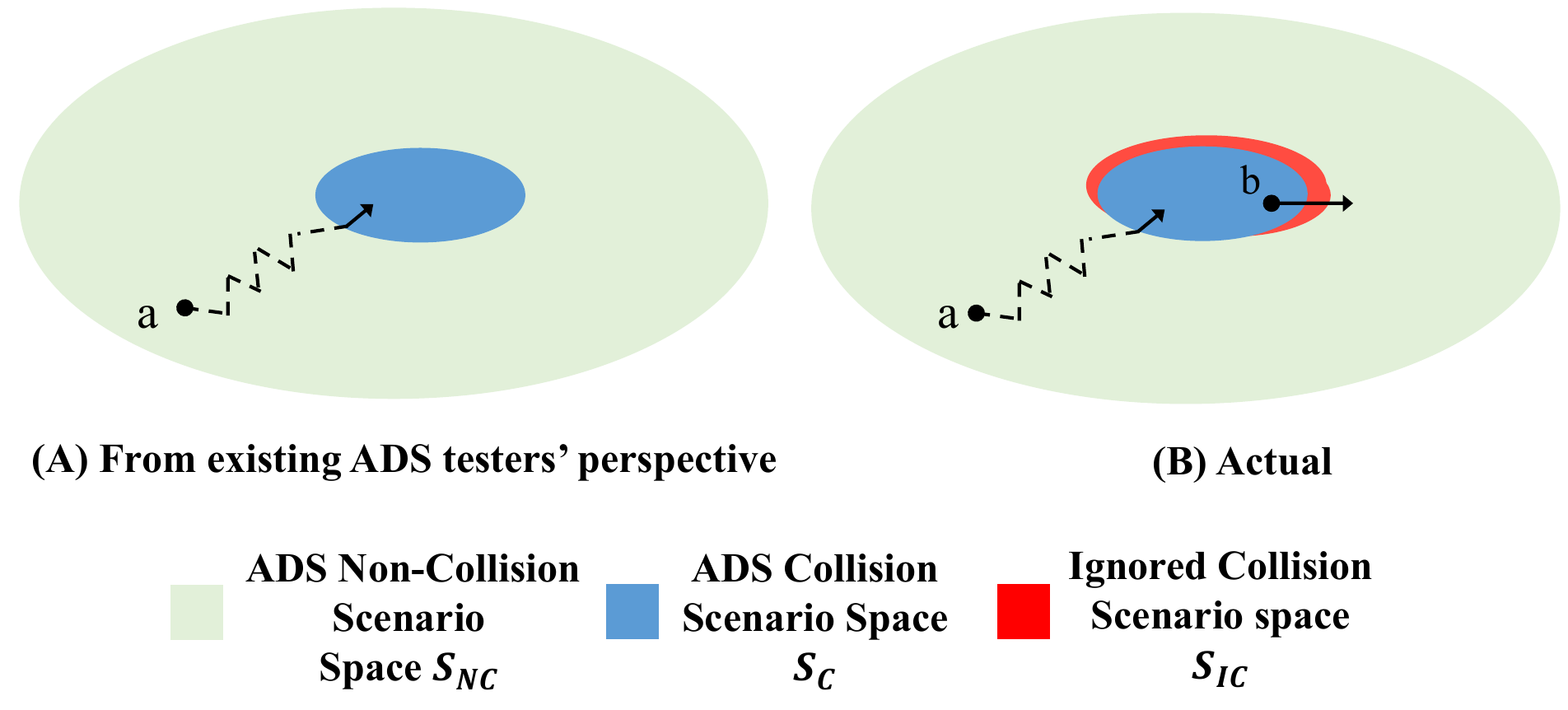} 
\caption{
ADS Scenario Space.
(A) The scenario space from the existing ADS testers' perspective, with only collision and non-collision scenarios.
(B) The actual scenario space, where the actual collision scenario space contains a ignored scenario space overlooked by ADS testers.
a: Existing ADS collision testing methods.
b: \sysname.
}

\label{fig:search_space}
\vspace{-6mm}
\end{figure}

To the best of our knowledge, no existing work considers the existence of ignored collisions in driving simulators, let alone identifies the scenarios causing ignored collisions. This paper aims to develop a systematic approach to identifying scenarios causing ignored collisions, referred to as \emph{ignored collision scenarios} (ICSs). To facilitate the understanding of ICSs and the illustration of our insights, we use Figure~\ref{fig:search_space} to illustrate the scenario space where each point in the scenario space is a driving scenario, which consists of the behaviors of all traffic participants and the environmental elements in a pre-defined time period. From the perspective of existing ADS testers, a point in the collision scenario space indicates that the subject vehicle collides with other traffic participants in this scenario according to the collision detector of the simulator, and vice versa for the non-collision scenario space as shown in Figure~\ref{fig:search_space} (A). Due to the imperfection of the collision detector in a simulator, the actual collision space includes ignored collision scenario space where the collision detector reports false negatives and, thus, is larger than the collision space discovered by the existing ADS testers, as shown in Figure~\ref{fig:search_space} (B). 
Our objective is to discover ICSs efficiently and delimit the actual collision scenario space to cover the ICSs thoroughly.

With no exception, the key to discovering ICSs is still to effectively generate inputs that lead to ignored collision scenario space.
One straightforward approach is to generate inputs from the non-collision inputs as existing efforts do to find collision scenarios ~\cite{drivefuzz}, i.e.,  method a in Figure~\ref{fig:search_space}.
Although such an approach may trigger some ICSs, 
the large search space involving a huge number of non-collision scenarios could make it deficient.
For example,
it can take up to 36 hours for Drivefuzz~\cite{drivefuzz} to detect collision scenarios while less than 0.03\% of them are ignored collision scenarios according to our evaluation in Section~\ref{sec:eval}.
Such ample irrelevant search space and ineffective guidance make it impractical to directly adapt the existing efforts on collision detection to search ICSs.

ICSs are fundamentally a sub-type of actual collision scenarios. By nature, ICSs share features similar to those of collision scenarios discovered by existing ADS testers. Slightly mutating those collision scenarios is more likely to transform them into ICSs than non-collision scenarios. This observation inspires us to generate inputs directly from the collision scenario space discovered by the existing ADS testers to improve the efficiency of discovering ICSs, as method b shown in Figure~\ref{fig:search_space} (B). For the simplicity of presentation and without ambiguity, the collision scenario space in the rest of the paper refers to the collision scenario space discovered by the existing ADS testers. 

To efficiently generate inputs from the collision scenario space, the following core challenges need to be addressed: 
\textbf{(1) Where to start in the collision scenario space?} 
\textbf{(2) How to guide ICS discovery? } 
To answer the questions above, we have empirically studied the real-world collision datasets and reports to identify the factors contributing to collisions, such as struck object type and collision speed. 
Specifically, based on the factors studied, we select representative collision scenarios that align with high-frequency collisions, mainly including but not limited to rear-end collisions.
Then, we manually determine the representative driving behaviors within each type of identified collision scenario, serving as the starting point of fuzzing.
To effectively generate ICSs, we propose a selective mutation strategy to 
slightly change representative scenarios while remaining in collision so that the small IC scenario space is less likely to be ignored.
Specifically, our mutation strategy adjusts the studied factors in the scenarios with adaptable step sizes toward the ICSs.

In this paper, we propose \sysname, a black-box fuzzing approach to efficiently discover the ICSs.
To the best of our knowledge, \sysname is the first work focusing on discovering the bugs of autonomous driving simulators.
Despite the high complexity of simulators, our experiment results demonstrate the effectiveness of our black-box method.
We implement \sysname on Carla~\cite{carla}, a widely used Unreal Engine-based driving simulator, and compare it with DriveFuzz~\cite{drivefuzz}, an SOTA simulation-based ADS fuzzer. 
The results reveal that \sysname outperforms DriveFuzz by finding 10\textasciitilde20x more ICSs with a 20\textasciitilde70x speedup. Such an efficiency improvement further enables \sysname to find seven more types of ICSs that DriveFuzz did not find.
Furthermore, \sysname can detect ~470 more ICSs compared to the state-of-the-art efforts, underscoring the effectiveness of our studied collision-contributing factors.
\sysname also demonstrates a significant real-world impact that the developer has confirmed all the discovered ICSs with one CVE ID assigned in the newest version of the most widely-used simulators, Carla. 


\noindent
In summary, this paper makes the following contributions: 
\begin{itemize} 
    \item We are the first to study the reliability of autonomous driving simulators and discover the existence of ignored collision scenarios.
    \item We identify the key factors that distinguish the ICSs from collision scenarios to narrow down the search space.
    \item We design a directed input generation method to effectively generate test inputs toward ICSs by selectively mutating the ICS-relevant collision factors stepwise.
    \item We provide empirical evidence to demonstrate that our approach significantly outperforms the SOTA simulation-based ADS fuzzer in identifying real-world ICSs.
    Bug records, demonstration videos, and supplementary experimental results can be found at \url{https://aoooooa.github.io/ICSFuzz_.github.io/}. 

\end{itemize}


\vspace{-2mm}
\section{Background \& Problem Definition}
\label{sec:background}
To better understand what ignored collision scenarios are and why existing efforts are deficient in discovering them, 
we first introduce the simulator's structure and where the ICSs can happen(\S~\ref{sec:background-simulator}).
Then, we outline the drawbacks when applying existing ADS testers to discover ICSs and our intuition to discover them efficiently (\S~\ref{sec:background-problem and motivation}).








\vspace{-1.5mm}
\subsection{Preliminary of AD Simulators}
\label{sec:background-simulator}
\vspace{-1mm}

\begin{figure}[t]
\centering
\includegraphics[width=1.0\linewidth]{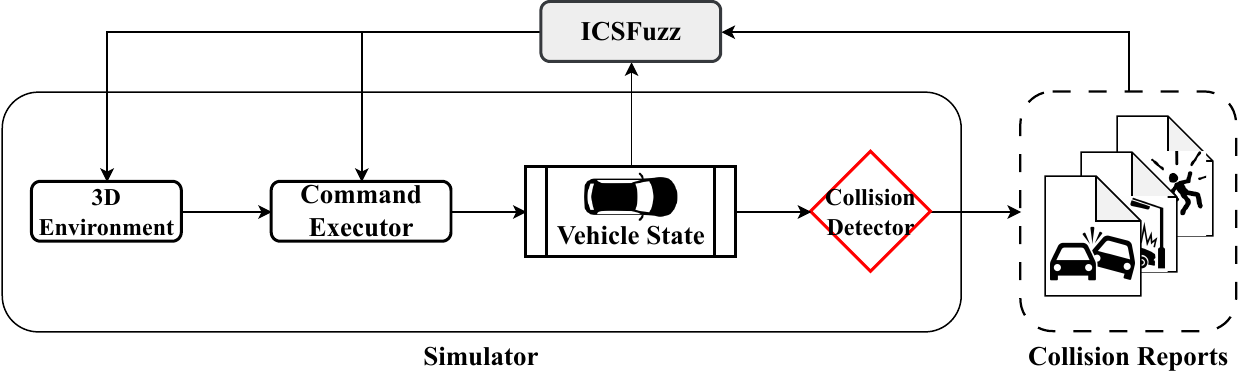} 
\caption{
Structure of AD simulator.} 
\label{fig:bg:simulator}
\vspace{-7mm}
\end{figure}

Inspired by the powerful game engine, simulators emulate a photo-realistic virtual environment that resembles the real world to provide an ideal platform for developing and testing autonomous vehicles.
The structure of a simulator is shown in Figure~\ref{fig:bg:simulator}.
ADS testers generate diverse 3D scenarios with calculated scenario configurations to interact with the simulator.
Eventually, the simulator operates the vehicle using the ADS under test in the generated 3D scenario to check whether a collision exists.
Due to the complexity of the simulators and the characteristics of interacting with the game engine, most of the testers adopt the black-box testing method and gain favorable results~\cite{drivefuzz,avfuzzer,avchecker}.

\textbf{Collision Detector}, as an essential module in the simulator, is responsible for monitoring and notifying collisions during the operation of the attached vehicle.
The detector helps identify and analyze critical scenarios that can aid in refining and improving the ADS's performance and safety measures.
The reliability of the collision detector is crucial as it directly affects our ability to identify accident scenarios accurately.

\textbf{Ignored Collisions} is defined as the false negative of the collision detector to report the collision.
As Figure~\ref{fig:motivating example} (C) shows, the right corner of the vehicle collides with the rear wheel of a bicycle, causing the bicycle to topple over. 
However, the failure of the collision detector means neither the vehicle nor the bicycle can report the collision,
which can cause lethal damage to the bicycle driver.
On the other hand, Figure~\ref{fig:motivating example} (B) portrays a collision that was successfully detected and reported to prevent further damage in time.
One crucial observation is that the factors causing the detection failure are minor since scenarios B and C can be almost identical regarding vehicle position and speed.  

\vspace{-1.5mm}

\subsection{Problem and Motivation}
\vspace{-1mm}

\label{sec:background-problem and motivation}

The existence of the ignored collision scenarios inspires us to reframe the problem from a new perspective, different from the assumption used in existing work.
Specifically, to better illustrate the relationship between ignored collisions and other scenarios, we denote all possible scenarios a running vehicle in the simulator can encounter as scenario space 
$\mathcal{S}$ in Figure~\ref{fig:search_space} (B) considering the presence of the ignored collision scenarios.
Each point in $\mathcal{S}$ represents a possible scenario. 
The entire scenario space involves three subspaces: non-collision scenario space $\mathcal{S}_{NC}$, collision scenario space $\mathcal{S}_{C}$, and ignored collision scenario space $\mathcal{S}_{IC}$, 
which implies $\mathcal{S} = \mathcal{S}_{NC} \cup \mathcal{S}_{C} \cup \mathcal{S}_{IC}$.

\subsubsection{\textbf{Existing simulation-based ADS Testing}}
\label{sec:bg:existing}
Existing simulation-based ADS testers typically initiate testing from non-collision scenarios, employing random mutation methods like genetic algorithms to generate scenarios potentially violating the ADS.
The testing round concludes upon encountering a violation.
While existing works have made significant efforts to test ADS using the simulator, they cannot be directly applied to ICS discovery due to ineffective input generation under different views of the problem.

\textbf{Irrelevant Input Generation with Overly-simplified Assumptions.}
Existing simulation-based testers solely focus on finding the collision scenarios resulting from the ADS failure. 
The assumption of the reliable simulator simplifies their view of the scenario search space as containing the collision scenarios  ($\mathcal{S}_{C}$) and non-collision scenarios ($\mathcal{S}_{NC}$), as shown in Figure~\ref{fig:search_space}(A). 
Simulation-based testers initiate their testing process with non-collision scenarios and iteratively generate new scenarios by mutating the existing ones guided by the execution feedback to identify events where the ADS fails~\cite{mosat,avchecker,avfuzzer}. 
Intuitively, their testing processes are illustrated as method a in Figure~\ref{fig:search_space} (B), where the search paths start from $\mathcal{S}_{NC}$ and proceed towards $\mathcal{S}_{C}$. 
The points along these search paths represent the test inputs generated during testing iterations.
However, as ICS is a previously-unaware type of collision, the actual collision space contains not only collision scenarios caused by ADS failure but also ignored collision scenarios (denoted by $\mathcal{S}_{IC}$),
as shown in Figure~\ref{fig:search_space}(B).
Thus, starting testing from the non-collision space is deficient, with explosive irrelevant inputs generated before detecting ICSs.


\textbf{Deficient Mutation Strategy with Unrelated Collision Factors.}
Except for the inefficient initial test cases, the input generation strategy adopted by existing simulation-based testers further decreases the efficiency of ICS discovery.
When generating a new test input, existing efforts~\cite{drivefuzz,wachi-ads-rl,avfuzzer} randomly mutate the test scenario with different factors, e.g., weather and light, based on the tested feedback, e.g., the number of accelerations to estimate the collision probability~\cite{drivefuzz}. 
Intuitively, randomness helps existing efforts identify ADS failures, as they can thoroughly explore the ADS collision space from different directions originating from the relatively ample non-collision space.
However, as ICSs only account for a minor proportion of collisions, the same intuition of randomly generating the input is no longer effective in detecting them.
These practices can overlook the small regions of $\mathcal{S}_{IC}$.
For example, randomly generating the inputs can cause the different inter-point distances and orientation changes along the zigzag search paths in method a, shown in Figure ~\ref{fig:search_space} (B).

\begin{figure}[t]
\centering
\includegraphics[width=1.0\linewidth]{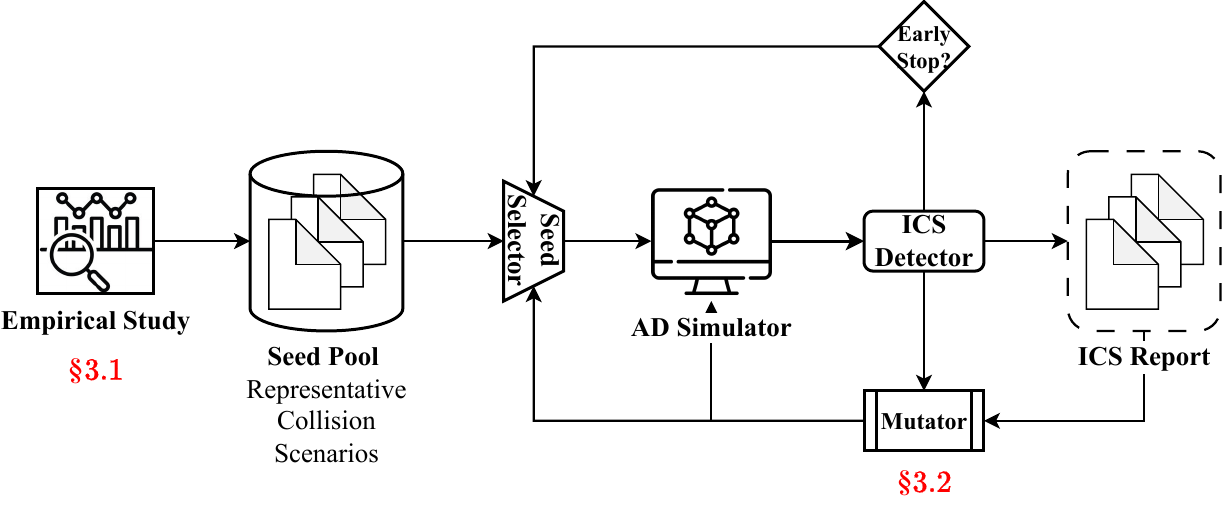}
\caption{Overview of \sysname. 
}
\label{figure:system_structure}
\vspace{-8mm}
\end{figure}

\subsubsection{\textbf{Our Observations and Insights}}
\label{sec:bg:our}
Our intuition to effectively discover ICSs comes with a novel view of their correlation with other scenarios in the simulator.
Our observation is that since ICSs are the false negative results of the collision detector, which an ICS must be a collision scenario, they only exist in the $\mathcal{S}_{C}$.
Thus, slightly mutating the collision scenarios is more likely to transfer them into ICSs than non-collision scenarios.
Based on these observations, our idea is to directly conduct the ICS discovery starting from  $\mathcal{S}_{C}$ with a selective mutation strategy targeting ICSs in a step-wise manner.
Moreover, if we can only mutate the related factors that lead to ICSs,
the effectiveness of input generation can be significantly improved with fewer irrelevant inputs generated.
Ideally, the search path can be shown as method b in Figure~\ref{fig:search_space}(B).
We generate collision scenarios that can be identified by the collision detector as test inputs and search toward $\mathcal{S}_{IC}$.
We save the useless search process within $\mathcal{S}_{NC}$ and traverse the $\mathcal{S}_{IC}$ effectively compared to method a.


Based on this intuition, we identify two main challenges for finding suitable starting points and mutating collision scenarios while discovering ICSs:

\textbf{C1. How to effectively find suitable starting points for detecting ICS?}
To search ICSs from $\mathcal{S}_{C}$, it is crucial to select appropriate starting points.
However, the randomly generated collision scenarios may result in repetitive ICSs with high false negative rates.
Hence, it is crucial to identify the representative collision scenario types as the foundation for selecting the starting points.


\textbf{C2. How to effectively mutate collision scenarios for detecting ICS?}
To effectively search ICSs, it is crucial to employ a mutation strategy that can effectively distinguish the minor difference between identified collision and IC.
Otherwise, we risk overlooking ICSs closely related to the collision scenarios.
Hence, it is crucial to accurately identify the collision-contributing factors and determine the direction toward ICSs, along with adaptable step sizes for these factors.

\begin{algorithm}[tp]
\caption{ICSFuzz Testing Procedure 
}
\label{algo_whole}
\begin{algorithmic}[1]
\Require Seed Scenarios
$S_{\scaleto{C\mathstrut}{5pt}}$,
Control parameters $parameters$, 
Early stop threshold $T_{\scaleto{NC\mathstrut}{5pt}}$
\Ensure ICSs $S_{\scaleto{ICS\mathstrut}{5pt}}$  
\State $S_{ICS} \gets \emptyset$
\ForEach {$scenario \in \mathcal S_{\scaleto{C\mathstrut}{5pt}} $} 
\Comment{\S~\ref{sec:overview:answer_challenge1}}
\label{algo1:select_seed_scenario}
    \State $Count_{\scaleto{NC\mathstrut}{5pt}} \gets 0$ \label{algo1:count_ic_to0}
    \For {$parameter$ \textbf{in} $parameters$} \label{algo1:for_parameter}
    \State $scenarios'$ $\gets$ \Call{Mutate} {$scenario$, $parameter$} \label{algo1:mutate}
    \Comment{\S~\ref{sec:overview:answer_challenge2}}
    \For {$scenario'$ \textbf{in} $scenarios'$}
        \State $state$ $\gets$ \Call{Simulate}{$scenario'$} \label{algo1:simulate}
        \State $scenarioType$ $\gets$ \Call{CheckIC}{state} \label{algo1:checkic}
        \If{$scenarioType == ICS$} 
            \State $S_{\scaleto{ICS\mathstrut}{5pt}} \gets S_{\scaleto{ICS\mathstrut}{5pt}} \cup \{scenario'\}$  
        \ElsIf {$scenarioType == NC$}  \label{algo1:determine_nc}
            \State $Count_{\scaleto{NC\mathstrut}{5pt}} \gets Count_{\scaleto{NC\mathstrut}{5pt}} + 1$ 
            \If{$Count_{\scaleto{NC\mathstrut}{5pt}} > T_{\scaleto{NC\mathstrut}{5pt}}$ } 
                \State $Count_{\scaleto{NC\mathstrut}{5pt}} \gets 0$
                \State \textbf{continue} \label{algo1:stop_current_round}
            \EndIf
        \EndIf
    \EndFor
    \EndFor
\EndFor
\end{algorithmic}
\end{algorithm}
\setlength{\textfloatsep}{1mm}

\vspace{-4mm}
\section{Overview of \sysname}
\label{sec:overview}
\vspace{-2mm}


Based on the observations in \S \ref{sec:background-problem and motivation}, we present \sysname, a novel black-box fuzzing approach to effectively and efficiently discover ICSs in the simulator.
As illustrated in Figure~\ref{figure:system_structure}, we begin from the determined collision scenarios and generate new undetermined collision scenarios to discover ICSs based on empirically studied collision contributing factors during the offline preparation stage. 
To effectively find suitable test starting points (\textbf{C1}), we identify the determined collision scenarios building upon the high-frequency collision scenario types like car following based on the empirically studied conclusions (\S~\ref{sec:overview:answer_challenge1} and \S~\ref{sec:design:offline}).
To effectively mutate undetermined collision scenarios for ICS discovery (\textbf{C2}), we propose a selective, step-wise mutation strategy to generate inputs that target ICSs directly based on the studied directions and experimentally determined step sizes toward the ICSs.

The testing procedure is described in Algorithm~\ref{algo_whole}, designed explicitly for ICS discovery.

\textbf{Initialization (Line \ref{algo1:select_seed_scenario}-\ref{algo1:count_ic_to0})}. 
Each testing round begins from a seed scenario, a determined collision scenario from the set of representative collision scenarios. 
The scenario includes environment settings(e.g. map, road structure) and driving behavior of traffic participants.

\textbf{New Inputs Generation (Line \ref{algo1:for_parameter}-\ref{algo1:mutate})}. 
We mutate the seed scenario according to the procedure in \S~\ref{sec:overview:answer_challenge2}, altering the driving behavior of traffic participants while keeping the environment constant to generate new collision scenarios. 
Specifically, we adjust control parameters related to ICSs based on collision contributing factors, using empirically determined directions and step sizes(\S~\ref{sec:emp_study}).

\textbf{Testing Round Termination (Line \ref{algo1:determine_nc}-\ref{algo1:stop_current_round})}. 
Unlike most testers, our approach continues testing after identifying an ICS.
We stop only after encountering a predetermined number of non-collision scenarios, indicating we have left the $\mathcal{S}_{C}$ and can stop the current round.
This procedure identifies diverse ICSs and provides insights into collision detector failure by comparing behavior control parameters.
\ignore{
Unlike most existing testers, our approach does not immediately terminate the current testing round and start a new one after identifying an ICS.
Instead, we continue the search process along the directions until a specific number of non-collision scenarios are encountered, indicating that we leave the collision scenario space and stop the current testing round.
By employing this testing procedure, we can not only discover diverse ICSs but also gain insights into the reasons for the failure of the collision detector by comparing the value of their behavior control parameters.}

\textbf{Monitoring (Line \ref{algo1:simulate}-\ref{algo1:checkic})}. During testing, we adopt the overlap between the ground-truth 3D object detection bounding boxes of the tested vehicle (referred to as ego vehicle (EV)) and the struck object (referred to as the none-player-character (NPC)) to determine the occurrence of ICSs, as listed in Algorithm~\ref{algo:algo_detector}. 
The essential characteristic of collision is physical contact, which can be approximated using the bounding boxes.

\begin{algorithm}[tp]
\caption{Ignored Collision Checker}
\label{algo:algo_detector}

\begin{algorithmic}[1]
\Require 
Bounding Box of Ego Vehicle $bbox_{ev}$,
Bounding Box of NPC $bbox_{npc}$,
Bounding Box Overlapping Threshold $T_{bbox}$, 
Simulator Built-in Collision Detector \Call{CD}{state}
\Ensure ScenarioType
\Function{CheckIC}{$bbox_{ev}$, $bbox_{npc}$, $T_{bbox}$, $state$}
\State $cond1$ $\gets$ \Call{IoU3D}{$bbox_{ev}$, $bbox_{npc}$} $\geq$ $T_{bbox}$
\State $cond2$ $\gets$ \Call{CD}{state}
\If {$cond1 == True \land cond2 == False$}
    \State \Return {IC}
\ElsIf {$cond1 == False \land cond2 == False$}
    \State \Return {NC}
\EndIf
\EndFunction
\end{algorithmic}
\end{algorithm}

\vspace{-1.5mm}

\subsection{Intuition 1: Influential Scenario Collections via High-Frequency Collisions}
\vspace{-1mm}
\label{sec:overview:answer_challenge1}

Intuitively, searching ICS from the small $\mathcal{S}_{C}$ is more efficient than the ample $\mathcal{S}_{NC}$.
However, unsuitable starting points within $\mathcal{S}_{C}$ may result in repetitive ICSs with high false negative rates.
Therefore, it is essential to use representative collision scenarios as starting points, ensuring they are relatively distant from each other to enhance the coverage of the space and improve the search for ICSs.


To address this, we conduct an empirical study on key factors causing real-world collisions.
Based on the studied collision-contributing factors, we identify high-frequency collisions, like rear-end collisions, that consistently appear in research studies.
Then, we determine representative scenarios, such as car following, which are prone to frequent collisions.
These scenarios serve as crucial starting points for our analysis.
For instance, in a car-following scenario, if the EV collides with the NPC vehicle in front, the EV should travel straight at 20m/s, while the NPC vehicle should move at 10m/s. This setup creates a determined collision, serving as an effective starting point.

\textbf{Remark.} By identifying representative collision scenarios as our starting points for testing, we achieve more comprehensive coverage of the space and enhance the effectiveness of discovering the ignored collision scenarios.

\begin{figure}[t]
\centering
\includegraphics[width=\linewidth]{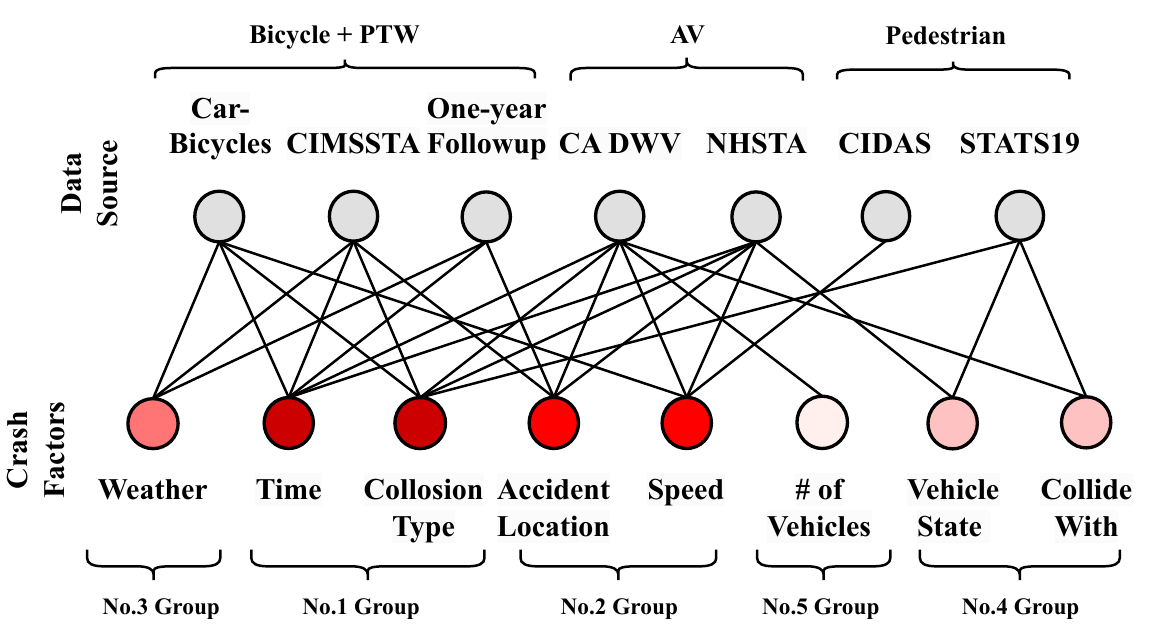}
\caption{
High-priority control parameters for collisions. Data source details are provided on our website.
}
\label{figure:CrashFactors_illustration}
\vspace{-1mm}
\end{figure}

\vspace{-1mm}
\subsection{Intuition 2: Selective Scenario Mutation with Collision-Contributing Factors}
\vspace{-1mm}

\label{sec:overview:answer_challenge2}
Due to the shared nature of ICSs with identified collisions, mutating collision-contributing factors can effectively generate new inputs toward ICSs. 
Since both involve collisions, selectively and slightly mutating the identified collision scenarios in an adaptable step-wise manner is more likely to transform them into ICSs than non-collision scenarios.
Hence, it is essential to identify key collision contributing factors within representative scenarios and determine the related behavior control parameters for traffic participants. 




Specifically, we adopt collision angle, speed, and distance as the control parameters to change the driving behaviors of EV and NPC for generating new test scenarios (the reason and the physical meaning of the parameters are described in \S~\ref{sec:design:offline}).
After determining the representative collision scenario, we sequentially modify its control parameters with adapted step sizes to generate new test scenarios. 
For example, suppose the collision angle of the last tested scenario is 0.02, and the search direction is to increase the collision angle with the step size of 0.02. The new test scenario with a collision angle of 0.04 is generated and sent to the simulator for testing. 
The increased direction and the step size, 0.02, are determined through empirical study ((\S~\ref{sec:ep:control_parameter}). 

\begin{table}[tp]
\centering
\caption{Meaning and dominant types of the collision contributing factors from our empirical studies.
}
\resizebox{\columnwidth}{!}{ 
\begin{tabular}{c|l|l|l}
\toprule
\# 
& \textbf{Factors}
& \textbf{Meaning}
& \textbf{Dominant Types(\%)}
\\ 
\midrule
01 
& Time 
& When the collision occurred. 
& \makecell[l]{
Daylight
(77.2\%~\cite{identifying_factors}, 67.3\%~\cite{avoid},
\\ 
64.52\%~\cite{analysis_poi}, 52.2\%~\cite{single_bicycle_crash}, 46.4\%~\cite{car-bicycle_analysis}, 
\\
45\%~\cite{automated_latent}), 
Dark (56.25\%)~\cite{divergent_effect}
}
\\
\hline
02 
& Weather 
& \makecell[l]{
The weather when \\ the collision occurred.
}
& \makecell[l]{
Clear weather (88.54\%~\cite{divergent_effect}, 
\\
82.8\%~\cite{single_bicycle_crash}, 77.42\%    ~\cite{analysis_poi}, 
\\ 
70.87\%~\cite{car-bicycle_analysis},
68.9\%~\cite{identifying_factors},
48\%~\cite{automated_latent})
}
\\
\hline
03
& Speed
& \makecell[l]{The vehicle's speed \\ when the collision occurred.} 
& \makecell[l]{ 
$\leq11.18 \text{m/s}$ (88.54\%)~\cite{divergent_effect}, 
\\
$\leq15 \text{m/s}$ (31.8\%)~\cite{avoid}, 
\\
$\leq10 \text{m/s}$ (50.8\%)~\cite{pedestrian_causation}
}
\\
\hline
04 
& \makecell[l]{Accident \\ Location}
& Where the collision occurred.
& \makecell[l]{
Intersection(73.5\%~\cite{exploratory_cal}, 69.72\%~\cite{learn_crash}, 
\\
65.63\%~\cite{divergent_effect}, 47.31\%~\cite{analysis_poi}),
\\
Highway(32\%~\cite{avoid}), 
\\
Straight Road(23.9\%~\cite{identifying_factors}),
\\
Urban(79\%~\cite{single_bicycle_crash})
}
\\
\hline
05 
& \makecell[l]{Vehicle State \\ While \\ Colliding}
& \makecell[l]{
The vehicle's specific driving 
\\
state or action when
\\
collision occurred.
}
& \makecell[l]{
Straight(87.5\%~\cite{divergent_effect}, 66.3\%~\cite{exploratory_cal})
\\
Stopped(36.08\%~\cite{analysis_poi})
}
\\
\hline
06 
& Collide With
& \makecell[l]{
The type of object \\ 
the vehicle collided with.
}
& \makecell[l]{
Car(90.88\%~\cite{typical_pedestrian}, 91.25\%~\cite{divergent_effect})}
\\
\hline
07
& 
\# of vehicles
& \makecell[l]{
The number of vehicles 
\\ 
Involved in the collision.
}
& \makecell[l]{
2 (90.3\%~\cite{exploratory_cal}, 87.5\%~\cite{divergent_effect}, 
\\
84.16\%~\cite{automated_latent})}
\\
\hline
08
& Collision Type
& \makecell[l]{
The relative positions and 
\\ 
movement patterns of 
\\ 
vehicles during collisions.
}
& \makecell[l]{
Rear-End(64\%~\cite{analysis_poi}, 61.1\%~\cite{exploratory_cal}, 
\\
59.38\%~\cite{divergent_effect},
49.85\%~\cite{learn_crash},
\\ 
46.9\%~\cite{avoid})
}
\\ 
\bottomrule
\end{tabular}
}
\label{tab:collision_factors_highest_type}
\vspace{-1.5mm}
\end{table}

\smallskip
\noindent
\textbf{Remark.}   By following the iterative approach in Algorithm~\ref{algo_whole}, we systematically analyze combinations of distance, speed, and angle values that affect the occurrence of ICS, enabling thorough simulator testing.
Moreover, such general factors can be adapted for arbitrary collision behavior in various scenarios, which provides better compatibility for validating the simulator's reliability and identifying the root causes of the failure in the simulator.

\vspace{-1mm}
\section{Effective ICS Generation}
\label{sec:design:offline}

To address \textbf{C1} and \textbf{C2}, we conduct an empirical study to determine the most appropriate factors for effectively discovering ICS.
We first collect papers and reports that record and analyze real-world collision events.
Then, we carefully identify the representative factors and their corresponding values, resulting in most collisions. 
As such, we can further design the representative collision scenarios, control parameters of the EV, and their respective search directions to generate new test inputs.
To optimize the ICS search efficiency, we experimentally determine the step sizes for control parameters.

\vspace{-1.5mm}
\subsection{Empirical Study}
\label{sec:emp_study}
\vspace{-1mm}

Although we mainly focus on collisions caused by autonomous vehicles, we still involve the accident data from human-driven vehicles due to the limited availability of autonomous driving data. Specifically, we focus on two types of data sources: academic papers from top journals and conferences and accident reports from government agencies. In terms of academic articles, we surveyed papers in the field of transportation for the past ten years, especially traffic accident studies.
Regarding the accident reports, we collect the national and regional authorities' published collision reports.
Eventually, we found $58$ papers related to the collision analysis and $2$ official collision reports from National Highway Traffic Safety Administration (NHSTA) and California Department of Motor Vehicles (CA DMV).

\begin{figure}[t]
\centering
\includegraphics[scale=0.45]{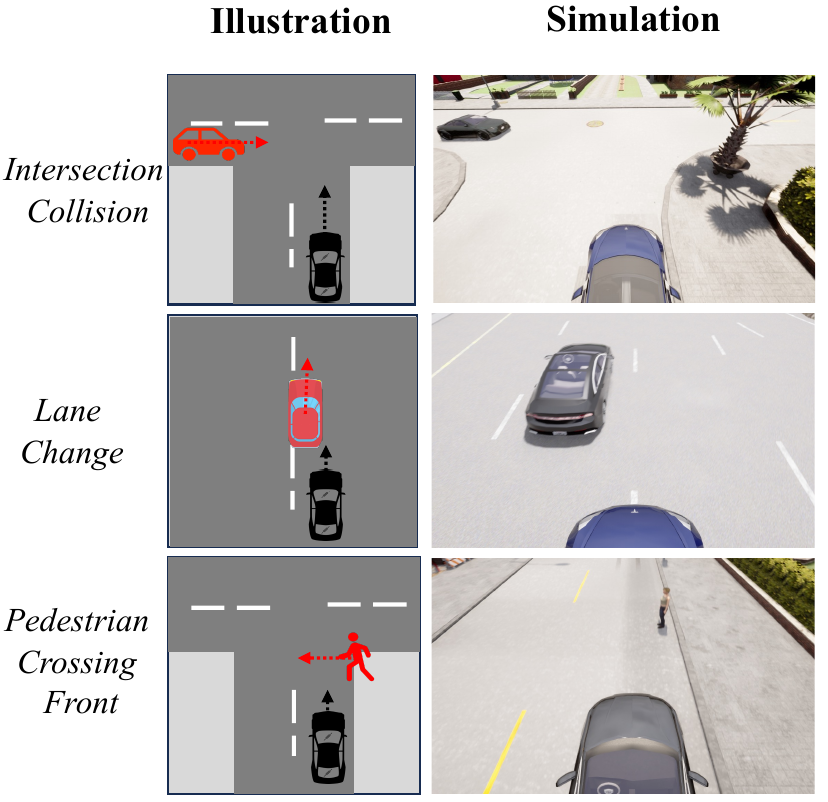}
\caption{Illustration of some collision scenarios. 
}
\label{fig:selected_scenarios}
\vspace{-1mm}
\end{figure}

We then distill the factors that contribute to collisions from the above works.
Some papers~\cite{divergent_effect,analysis_poi, exploratory_cal, learn_crash} utilize the NHSTA pre-crash collision topology~\cite{NHSTA_pre-collision} to describe their collected accident data, calculating and presenting ratios associated with the extracted factors when feasible. Furthermore, these papers incorporate combinations of factors to pursue their research objectives.
In contrast, other papers~\cite{learn_crash,identifying_factors,automated_latent} delve into crash reports and employ techniques such as clustering or natural language processing to extract impact factors and determine their occurrence frequency ratios.
Additionally, certain studies~\cite{avoid} attempt to create a novel collision dataset by combining data from various sources, including CA DMV~\cite{DMV_reports} and NHSTA~\cite{NHSTA_pre-collision} collision reports, collision data collected from social media and news, and a detailed description of the new dataset.

We have comprehensively summarized collision contributing factors, their values, and frequency from various studies. Detailed results are available on our website.
The simplified version can be found in Figure~\ref{figure:CrashFactors_illustration} and Table~\ref{tab:collision_factors_highest_type}, where we aggregated similar factors to emphasize their different levels of importance.
Notably, "Time" and "Collision Type" emerge as the most frequently discussed factors, followed by "Accident Location" and "Speed". 
"Weather" and "\# of Vehicles" are also frequently considered, followed by "Vehicle State While Colliding" and "Collide With".
"\# of Vehicles", "Accident Location", "Time" and "Location" are not considered in our ICS discovery.
Reasons are on our website. 
\ignore{

\begin{figure}[t]
	\begin{minipage}{0.32\linewidth}
		\vspace{3pt}
		\centerline{\includegraphics[width=\textwidth]{fig/ep_unrelated_factors/ep-ur_factors-road.png}}
	\end{minipage}
	\begin{minipage}{0.32\linewidth}
		\vspace{3pt}
		\centerline{\includegraphics[width=\textwidth]{fig/ep_unrelated_factors/ep-ur_factors-weather.png}}
	\end{minipage}
	\begin{minipage}{0.32\linewidth}
		\vspace{3pt}
		\centerline{\includegraphics[width=\textwidth]{fig/ep_unrelated_factors/ep-ur_factors-time.png}}
	\end{minipage}
	\caption{Analysis of Non-Contributing Factors in ICS discovery}
	\label{ figure:proportion-UnrelatedFactors}
\end{figure}
 
 }

\vspace{-1mm}
\subsection{High-Frequency Collisions Determination}
\label{sec:emp:scenario_selection}

Based on the aforementioned collision contributing factors, we select the high-frequency collision scenarios covering different values of "Collision Type" and "Collide With".
Different values of "Collision Type", such as rear-end and broadside, result in varying mechanical effects, object trajectories, and object deformation depth. 
Different values of "Collide With", such as vehicles and pedestrians, have different physical properties and structures, which result in different collision severity.
For example, collisions between vehicles may result in vehicle deformation, while collisions involving pedestrians may result in a pedestrian death.
Thus, collision scenarios involving varying "Collision Type" and "Collide With" can exhibit distinct collision patterns.
Such diversities ensure that the generated collision scenarios are diverse enough, thereby increasing the coverage of the collision scenario space.



\ignore{
According to our empirical study, the factor "\# of Vehicles" is not considered for scenario generation because of the difficulty of modeling multi-vehicle collision scenarios in the current simulators, and the two-vehicle collisions are the dominant scenarios.
The impact of "Accident Location" on collisions is primarily due to complicated road and traffic conditions, which may affect the reaction time for human drivers and the performance of the planning module for ADS.
Factors such as "Time" and "Weather" are significant in driving conditions as they also affect visibility and road surface conditions. 
These factors primarily impact the perception module of the ADS. 
However, it is essential to note that our main target is the reliability issues in the simulator rather than the ADS. Hence, the EV is intentionally not operated in fully autonomous mode and assumed to perform perfectly to facilitate the discovery of ICSs. Thus, the above four factors are not considered as the key factors in our cases. 
The experimental results demonstrating the irrelevance of unselected factors to the final ICS proportion in the generated test cases are provided on our website.}

Given the above key factors, we select "Lane Change," "Intersection Collision," and "Follow Leading Vehicle" as the three primary values of "Collision Type" to create the fundamental collision scenarios, which are prioritized in descending order of frequency: rear-end, sideswipe, and broadside. 
\textit{1) Follow Leading Vehicle (FLV):} The NPC vehicle is moving forward at x m/s; the ego vehicle typically follows from behind, resulting in rear-end collisions.
\textit{2) Lane Change (LC):} The initial setup of this scenario resembles that of "Follow Leading Vehicle", but the road structure of this scenario is on the wide highway for sideswipe collisions to occur. 
\textit{3) Intersection Collision (InC):} At T-intersections or four-way intersections, the NPC vehicle moves in a perpendicular direction to the ego vehicle as it proceeds forward, resulting in a broadside collision.

\begin{figure}[t]
\centering
\includegraphics[scale=0.4]{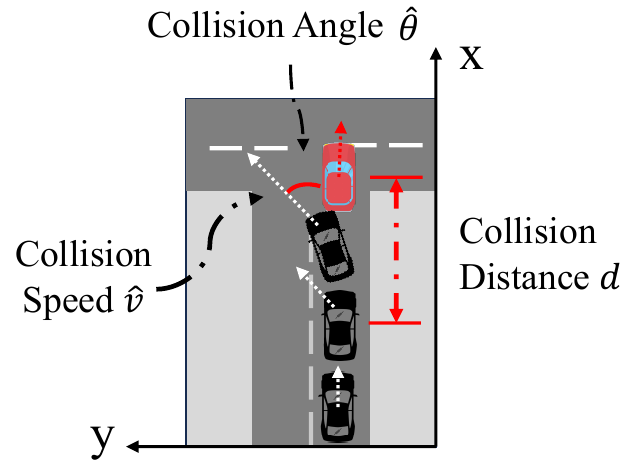}
\caption{
Illustration of adjustable control parameters for generating new inputs.
1) Collision Distance $d$. The distance between two vehicles when the EV alters its driving behavior.
2) Collision Speed $v$. The sustained driving speed of EV after changing driving behavior.
3) Collision Angle $\hat{\theta}$. The driving direction of EV after changing driving behavior. 
}
\label{figure:control_parameter}
\vspace{-1mm}
\end{figure}

As for the factor "Collide With", which encompasses collisions with various types of NPC actors as the struck object, we introduce three additional scenarios: "Follow Leading Bicycle", "Pedestrian Crossing Front", and "Pedestrian Standing Front." 
\textit{1) Follow Leading Bicycle (FLB):} 
The environment configuration of this scenario is the same as the "Follow Leading Vehicle".
The difference is that the NPC vehicle is replaced with an NPC bicycle. 
\textit{2) Pedestrian Crossing Front (PCF):}
In this scenario, the ego vehicle goes straight while the NPC pedestrian crosses the road.
\textit{3) Pedestrian Standing Front (PSF):} 
Some existing efforts~\cite{pedestrian_causation,typical_pedestrian} suggest that the stationary or movement of pedestrians can significantly impact the occurrence of collisions.
To examine the impact of the pedestrian's moving state on ICS occurrences, we introduce a new scenario called "Pedestrian Standing Front" for comparison with the "Pedestrian Crossing Front" scenario, where the NPC maintains a constant speed.
In the "Pedestrian Standing Front" scenario, the configuration remains identical to the "Pedestrian Crossing Front" scenario, with the only difference being that the pedestrian is static on the roadside. 





\vspace{-1mm}
\subsection{Control Parameters Selection}
\label{sec:ep:control_parameter}
\vspace{-1mm}

After selecting representative collision scenarios as the test starting points, the next step is identifying the specific control parameters with proper corresponding mutation directions and searching step sizes to mutate new test scenarios targeting ICSs.
Based on the intuition that ICSs are similar to identified collisions, we aim to slightly mutate the inputs so that the newly generated collision scenarios can be similar to those of their adjacent ones and, eventually, increase the likelihood of discovering ICSs.



\subsubsection{\textbf{Control Parameters and Search Directions}}

To determine the control parameters for mutation, we focus on "Speed" and "Vehicle State While colliding", which are the two most important ones, as shown in Figure~\ref{figure:CrashFactors_illustration}.  
Increasing "Speed" leads to higher impact forces during collisions, resulting in varying degrees of vehicle deformation and damage, referred to as \textit{collision speed}.
"Vehicle State While Colliding" denotes the actions of vehicles at the moment of collision, such as traveling straight or turning left. 
This factor comes from the relative direction of the vehicles at the moment of collision.
We refer to this relative direction as the \textit{collision angle}.
We also propose mutating the \textit{collision distance} factor, as smaller distances increase the likelihood of collisions.
The physical meaning of these three parameters is shown in Figure~\ref{figure:control_parameter}.


The search direction of these three control parameters to generate new scenarios under different collision scenarios identified in \S~\ref{sec:emp:scenario_selection} is elaborated on as follows. 

\textit{1) Collision Angle $\hat{\theta}$}. Different collision angles result in different values of "Vehicle State While colliding" and further result in different damage areas of the struck object within the specific "Collision Type."
For example, if the collision angle $\hat{\theta}$ in Figure~\ref{figure:control_parameter} is $0^\circ$, 
the EV will collide directly with the rear center of the NPC vehicle, 
which is a simple rear-end collision.
If the collision angle $\hat{\theta}$ is $45^\circ$, 
the EV will scrape against the left rear corner of the NPC vehicle, 
which can be treated as a specific sub-type of rear-end collision.


For the search direction of $\hat{\theta}$, we observed from Table~\ref{tab:collision_factors_highest_type} that the highest percentage of "Vehicle State While Colliding" involved with vehicles "going straight", followed by "left turn" or "right turn".
Based on the observation and our mutation strategy, we adopt the search direction that starts with the most frequent collision scenarios and targets the less frequent ones.
Therefore, our search direction for $\hat{\theta}$ begins at $0^\circ$ concerning the NPC and gradually changes to $90^\circ$ or $-90^\circ$.
\ignore{
For the search direction of $\hat{\theta}$, referring to the results presented in Table~\ref{tab:collision_factors_highest_type}, we observed that the highest percentage of "Vehicle State While Colliding" was associated with vehicles "going straight", followed by "left turn" or "right turn".
Based on this observation and our mutation strategy, we adopt the search direction that starts with the most frequent collision scenarios and targets the less frequent ones since we aim to mutate more collision scenarios and transform them into ICSs while searching toward the non-collision scenarios space.
Therefore, our search direction for $\hat{\theta}$ starts from a direction that is $0^\circ$ concerning the NPC, then gradually increases to $90^\circ$ or decreases to $-90^\circ$.
}

\textit{2) Collision Speed $\hat{v}$}:  
The findings presented in Table~\ref{tab:collision_factors_highest_type} highlight that collisions occur more frequently at lower speeds.
Thus, to mutate more collision scenarios, we should start from a lower collision speed and toward the higher values.
Based on the typically imposed speed limits on US interstate highways, which range from $65$ to $80$ mph, we begin the input generation process with a speed close to $0$ and gradually increase it toward the speed limit.
To explore speeds beyond the limit for ICS discovery, we set the maximum collision speed at 50m/s(111.85 mph), exceeding typical limits.


\textit{3) Collision Distance $d$:}
The search direction for collision distance starts from $2$m and increments up to $7$m, accounting for vehicle length(3.7-5m)~\cite{carlength}.
We measure the distance from the vehicle center rather than its boundary.
Without precise collision distance data, the search direction and starting point are chosen based on experience.
We hypothesize that altering driving behavior increases collision likelihood, so we explore distances from near to far.
The starting point at $2$ m is the minimum distance for collisions with smaller objects such as bicycles or pedestrians. The $7$m distance was set as the upper bound for potential collisions, with an additional $2-3$m buffer for exploration. Distances beyond this range offer no significant benefit.
The final results explored and analyzed the relationship between distance and ICS occurrences.

\subsubsection{\textbf{Search Step Size}}
\label{sec:emp-search_step}

\begin{figure}[tp]
  \begin{subfigure}[t]{0.23\textwidth}
    \includegraphics[width=\textwidth]{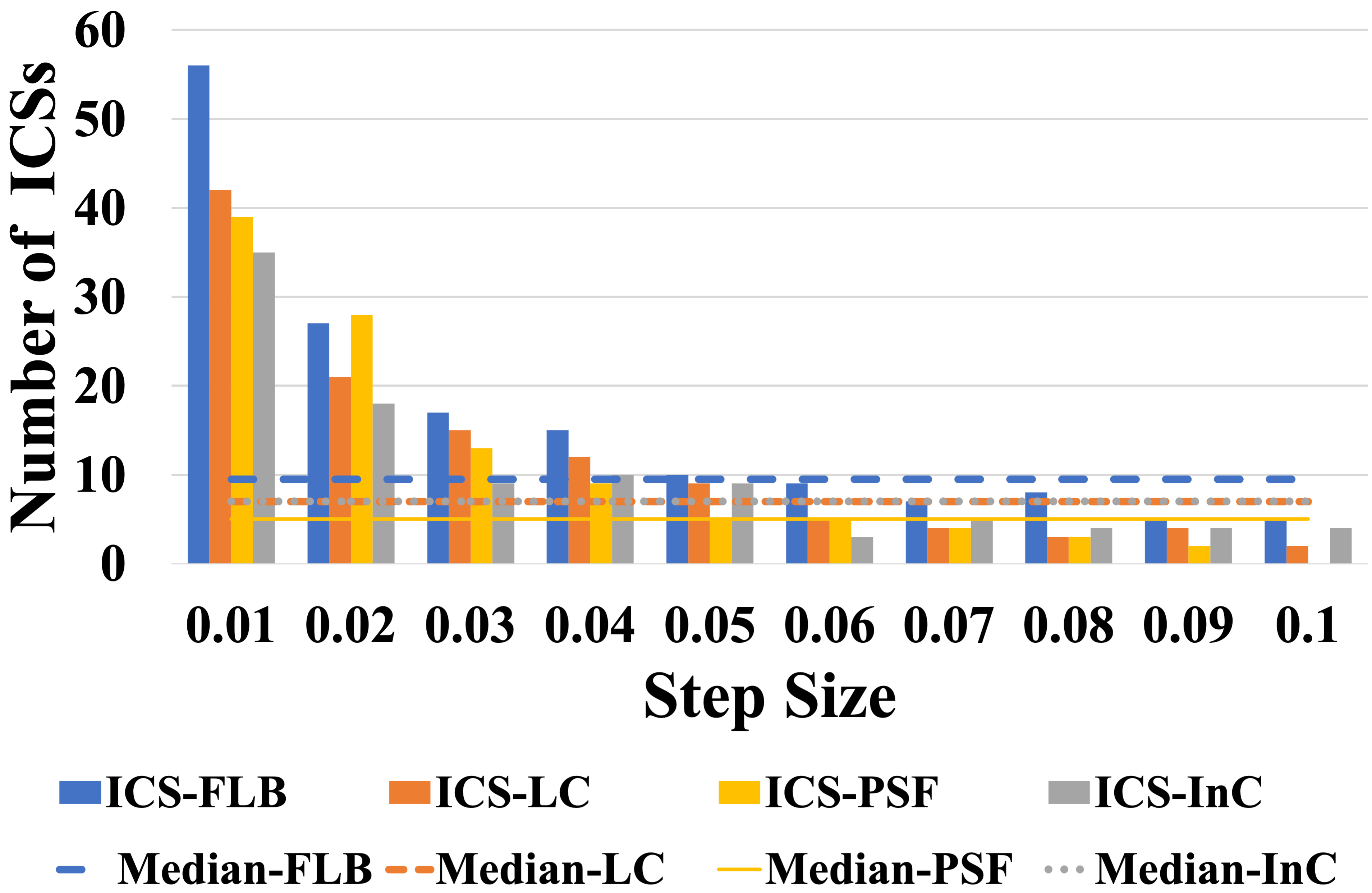}
    \caption{Angle x}
    \label{fig:parameter_x_flb}
  \end{subfigure}
  \hfill
  \begin{subfigure}[t]{0.23\textwidth}
    \includegraphics[width=\textwidth]{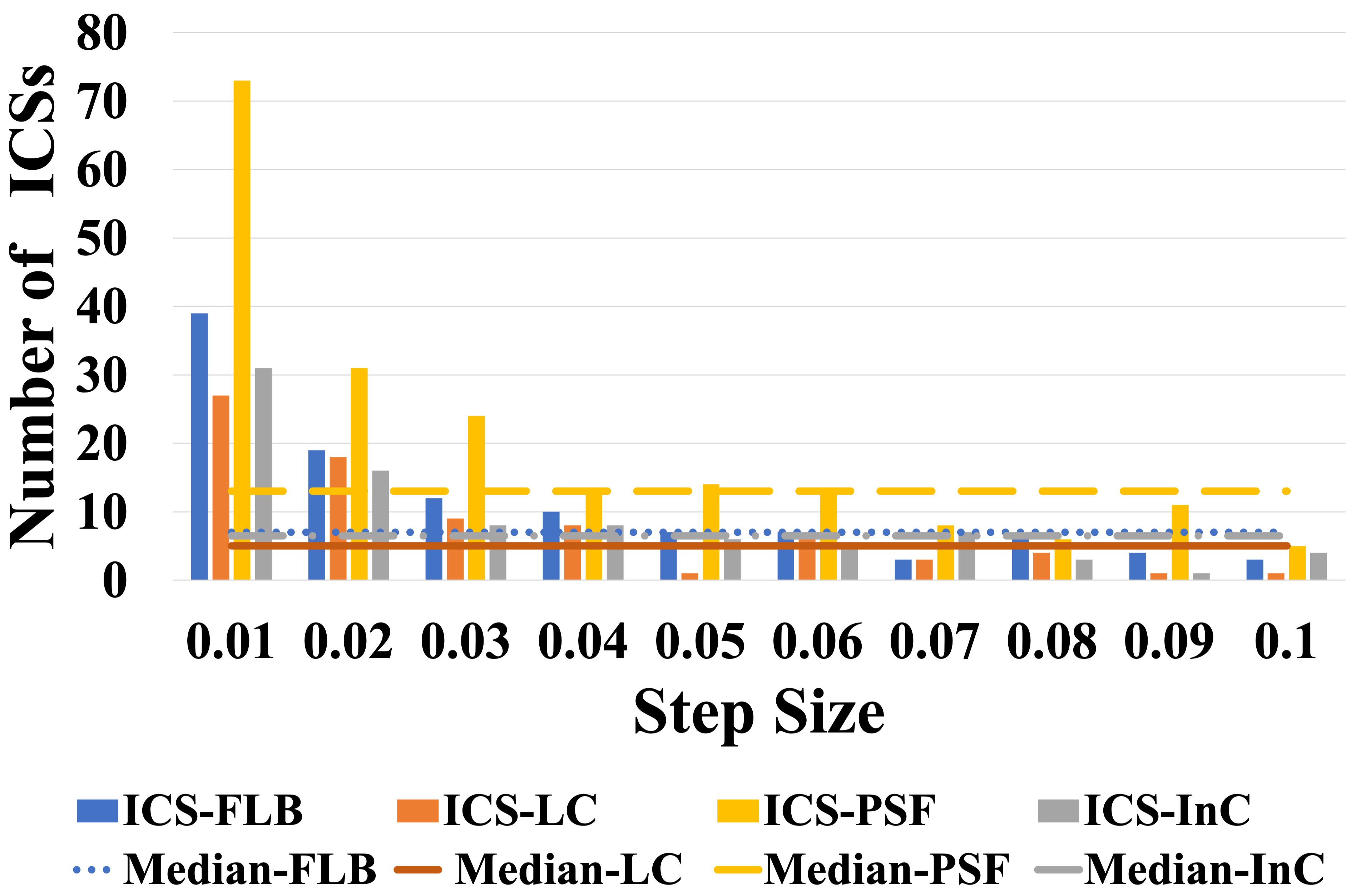}
    \caption{Angle y}
    \label{fig:parameter_y_flb}
  \end{subfigure} 
  \vspace{-1mm}
  \begin{subfigure}[t]{0.23\textwidth}
    \includegraphics[width=\textwidth]{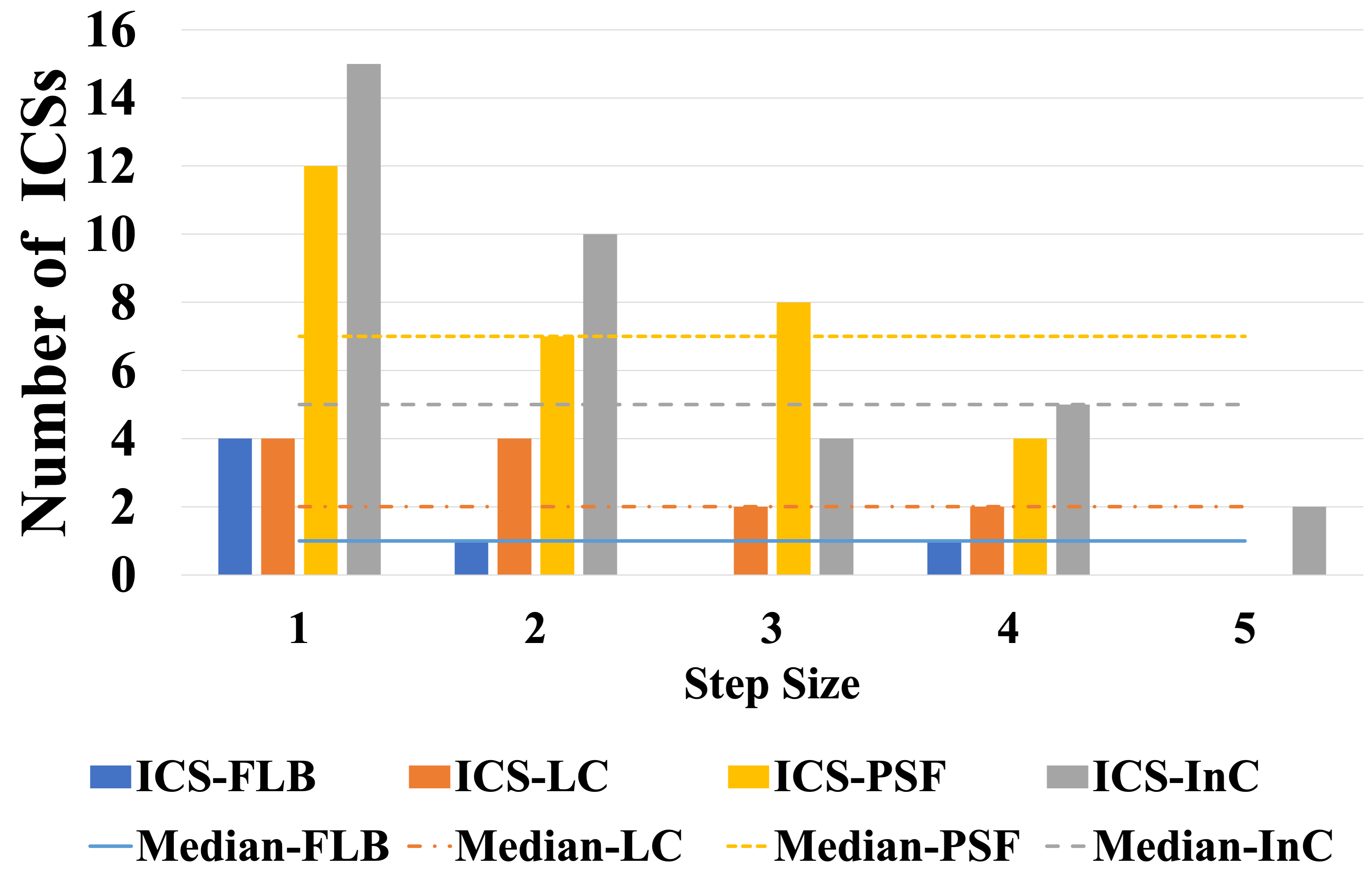}
    \caption{Distance}
    \label{fig:parameter_distance_flb}
  \end{subfigure}
  \hfill
  \begin{subfigure}[t]{0.23\textwidth}
    \includegraphics[width=\textwidth]{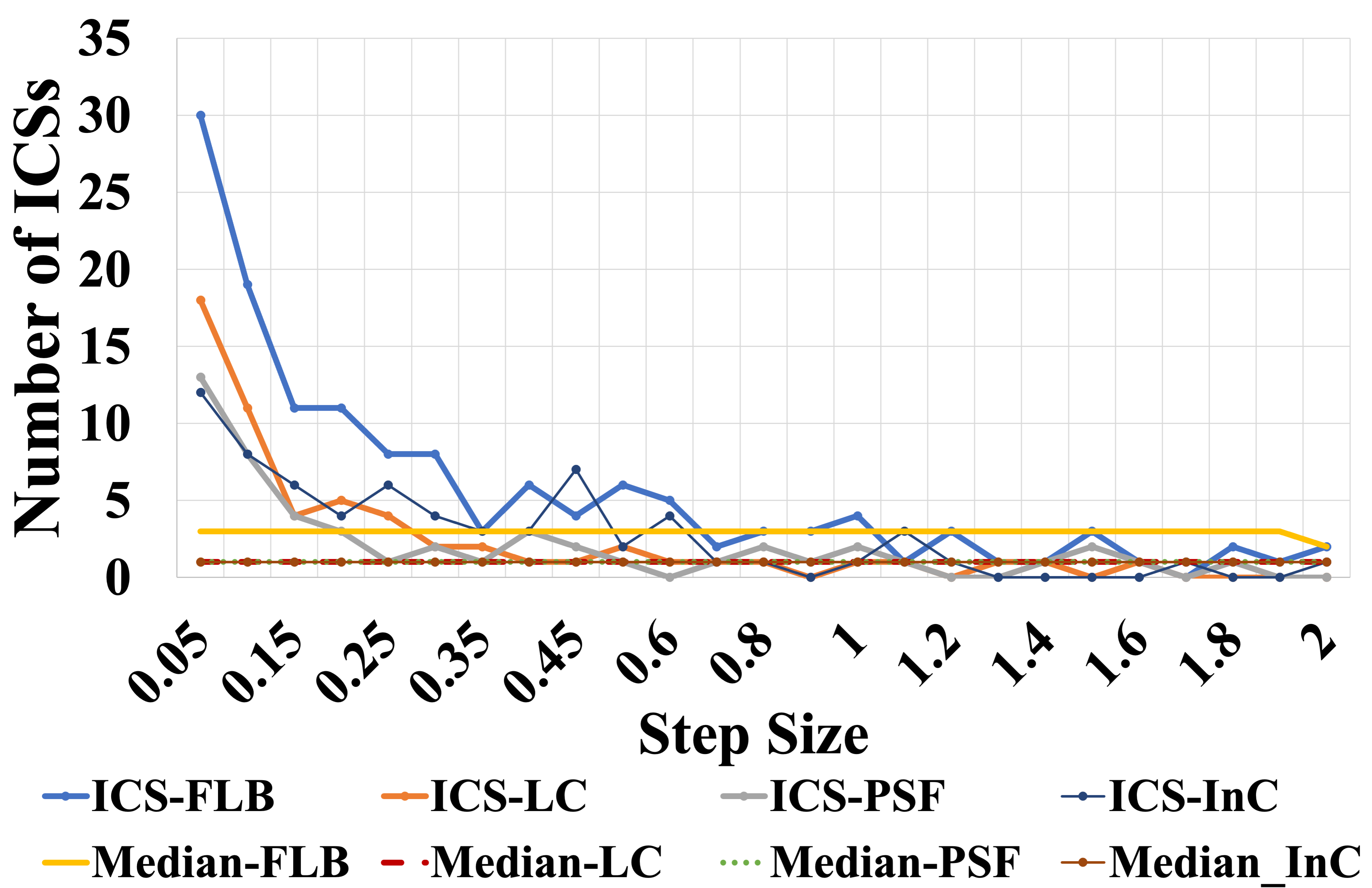}
    \caption{Speed}
    \label{fig:parameter_speed_flb}
  \end{subfigure}
  \vspace{-1mm}
\caption{The relationship between the number of ICSs and the control parameters' step sizes.}
\label{fig:step_size}
\vspace{-1mm}
\end{figure}

To determine the suitable search step size for control parameters when generating inputs, we empirically examine the relationship between the number of ICSs and step sizes for each control parameter in different collision scenarios.
We select the most appropriate sizes for $\hat{\theta}$, $d$, and $s$ within each type of collision scenario. 
The different control parameters are run in scenarios FLB, LC, PCS, and InC because they correspond to distinct collision areas of the struck object resulting from variations in environmental settings. 
A larger step size for $\hat{\theta}$ may result in missed collisions with bicycles but not with vehicles, due to the larger potential collision area of vehicles.
\ignore{For instance, when changing the collision angle from $0^\circ$ to $-45^\circ$ with a step size of $10^\circ$, there may be a continuous series of collisions observed in the LC scenario compared to the FLB scenario. This is due to the larger possible rear collision area provided by the volume of vehicles, as opposed to bicycles or pedestrians.}
Therefore, a uniform step size cannot be applied across all scenarios, and it is necessary to consider the step size separately for each scenario.

During testing, we keep other control parameters fixed while varying the tested parameter within its defined range. We increment or decrement the tested parameter using different step values and count the resulting ICSs. To ensure fairness, we randomly select values for the other control parameters and test the relationship between the final ICS count and the step size over ten rounds, averaging the results. We use the median as the threshold to help determine the exact step value since it can be a more representative metric to guarantee sufficient ICSs with less repetition in the final results. The results are shown in Figure~\ref{fig:step_size}. Notably, for collision angles in Carla, which involves a two-dimensional vector with $x$ and $y$ components, we conduct separate tests for each component.

\ignore{During testing, we maintain the other control parameters at fixed values while varying the tested control parameter within its defined or available range. 
We use different step values to increment or decrement the tested control parameter and count the resulting number of ICSs.
To ensure fairness in our analysis, we randomly select values for the other control parameters and test the relationship between the final ICS number and the test control parameters' step size for ten rounds.
We then calculate the average of the results. 
We use the median as the threshold to help determine the exact step value since it can be a more representative metric to guarantee sufficient ICSs with less repetition in the final results.
The outcome is presented in Figure~\ref{fig:step_size}.
It is worth noting that when it comes to the collision angle, the operation in Carla involves a two dimensional vector containing $x$-axis and $y$-axis components. Therefore, we conduct separate tests for different components.
}

For scenario FLB, we observe that when the search step size of the collision angle on the x-axis is smaller than $0.06$, the final ICS number is larger or equal to the median. 
Thus, we select the step size at $0.04$ to guarantee sufficient ICSs with less repetition.
The step size for the collision angle on the y-axis is $0.03$.
For collision distance, we choose the step size of $1$ as the steps at $2$ and $4$ match the median, risks missing ICS.
For collision speed, we select the step at $1$ since the step sizes larger than $2$ only result in $1$ ICS or even fewer, which is not ideal for ICS discovery.
We have determined the appropriate step sizes for different parameters in various scenarios.
Detailed values are available on our website.
\ignore{
Based on a similar process, for scenario LC, we have selected the following step sizes: $0.05$ for the angle at $x$-axis, $0.04$ for the angle at $y$-axis, $1$ for the collision distance, and $1$ for the speed step.
For the scenario PSF, we have selected the following step sizes: $0.03$ for the angle at $x$-axis, $0.03$ for the angle at $y$-axis, $1$ for the collision distance, and $1$ for the speed step.
For the scenario InC, we have selected the following step sizes: $0.05$ for the angle at $x$-axis, $0.02$ for the angle at $y$-axis, $4$ for the collision distance, and $1$ for the speed step.
The step size of collision distance for the scenario InC is because of the scenario specialties. 
To create the collision at the intersection, the distance between the initial location of the two vehicles is far since they are located at the ends of two vertical roads. 
If the EV were to change its driving behavior when it is close to NPC, it is highly possible to miss it since they are driving on perpendicular roads.
Thus, we select the step size of collision distance to $4$ and explore the relationship with the final ICS occurrences.
}

\begin{table*}[tp]
\centering
\caption{Results of ICS Detection by \sysname in Carla and Comparative Analysis with DriveFuzz, \sysnamenc, and \sysnamerandom. Time represents the average time taken to discover the specific categories of ICSs, while the proportion indicate the ratio of identified ICSs to the total generated tested scenarios.
}
\resizebox{\textwidth}{!}{
\begin{tabular}{c|c|ccc|cccc|c|ccc|cc|cc|cc}
\toprule
\multirow{3}{*}{
\textbf{No.}} 
& \multirow{3}{*}{\textbf{Scenario}} 
& \multicolumn{3}{c|}{\textbf{Control Parameters}}
& \multicolumn{4}{c|}{\textbf{Collision Factor}} 
& \multicolumn{1}{c|}{
    \multirow{3}{*}{
        \makecell{\textbf{Attack} \\ \textbf{Results}}
    }
} 
& \multicolumn{3}{c|}{\textbf{DriveFuzz}} 
& \multicolumn{2}{c|}{\textbf{ICSFuzz-NC}}
& \multicolumn{2}{c|}{\textbf{ICSFuzz-Random}}
& \multicolumn{2}{c}{\textbf{ICSFuzz}}
\\ 
& %
& \multicolumn{1}{c}{
    \begin{tabular}[c]{@{}c@{}}
    Collision \\ Distance
    \end{tabular}
} 
& \multicolumn{1}{c}{
    \begin{tabular}[c]{@{}c@{}}
    Collision \\ Speed
    \end{tabular}
} 
& \multicolumn{1}{c|}{
    \begin{tabular}[c]{@{}c@{}}
    Collision \\ Angle
    \end{tabular}
} 
& \multicolumn{1}{c}{
    \begin{tabular}[c]{@{}c@{}}
    NPC \\ Type
    \end{tabular}
} 
& \multicolumn{1}{c}{
    \begin{tabular}[c]{@{}c@{}}
    Road \\ Type
    \end{tabular}
} 
& \multicolumn{1}{c}{
    \begin{tabular}[c]{@{}c@{}}
    Collision \\ Position
    \end{tabular}
}
& \multicolumn{1}{c|}{
    \begin{tabular}[c]{@{}c@{}}
    Motion \\ Status
    \end{tabular}
} 
&  
& \# of ICS 
& Time 
& Proportion
& Time 
& Proportion
& Time
& Proportion
& Time

& Proportion

\\ 
\midrule
01 & FLB 
& L, M, F & L, M, H & N, P 
& \faBicycle 
& City road 
& \multicolumn{1}{c}{
    \begin{tabular}[c]{@{}c@{}}
    Corner \\ Collision
    \end{tabular}
} 
& Moving & G,K 
& - & - & -
& $\sim$0.34h & 2.32\%
& $\sim$0.11h & 2.25\% 

& $\sim$0.05h
& 40.95\%
\\
02 & FLB & F & L, M & 0 
& \faBicycle & City road & Rear end & Moving & F 
& - & - & -
& $\sim$0.6h & 5.81\%
& $\sim$0.4h & 4.93\%
& $\sim$0.02h
& 17.55\% 
\\ 
\midrule
03 & FLV & F & L, M, H & N, P  
& \faCar & City road & 
\multicolumn{1}{c}{
    \begin{tabular}[c]{@{}c@{}}Corner 
    \\
    Collision
    \end{tabular}
} 
& Moving 
& G 
& $\checkmark$ & $\sim$3.3h & 1.15\% 
& $\sim$0.31h & 9.13\%
& $\sim$0.14h & 8.16\%
& \textbf{$\sim$0.04h} &\textbf{25.6\%}  
\\ 
\midrule
04 & LC & F & L, M, H & N, P 
& \faCar & Highway 
& \multicolumn{1}{c}{
    \begin{tabular}[c]{@{}c@{}}Corner 
    \\ 
    Collision\end{tabular}
} 
& Moving 
& G
& - & - & - 
& $\sim$0.48h & 2.99\%
& $\sim$0.125h & 3.11\%
& $\sim$0.03h & 12.42\%
\\
05 & LC & M & M & N, P 
&  \faCar & Highway & Sideswipe & Moving & G 
& - & - & -
& $\sim$0.99h & 0.52\%
& $\sim$0.17h & 1.15\% 
& $\sim$0.03h &8.28\% 
\\ 
\midrule
06 & InC & L, M & L, M & P  
& \faCar & City road & Broadside & Moving & G 
& $\checkmark$ & $\sim$1.3h & 1.03\% 
& $\sim$0.34h & 2.1\%
& $\sim$0.23h & 1.05\%
& \textbf{$\sim$0.02h} & \textbf{11.46\%} 
\\
07 & InC & H & M, H & N  
& \faCar & City road & Broadside & Moving & G
& - & - & - 
& $\sim$0.72h & 3.01\%
& - & -
& $\sim$0.03h &7.64\%
\\ 
\midrule
08 & PSF & L & L, M & 0 
& \faUser & City road & Sideswipe & Static & G, K
& - & - & - 
& $\sim$0.63h & 4.51\%
& $\sim$0.15h & 3.74\%
& $\sim$0.02h &14.49\%
\\
09 & PSF & M, H & H & P 
& \faUser & City road & Rear end & Static & G 
& - & - & -
& $\sim$0.56h & 3.95\%
& $\sim$0.25h & 5.92\%
& $\sim$0.03h &9.56\%
\\ 
\midrule
10 & PCF & L, M & L, M, H & N, P  
& \faUser & City road & Sideswipe & Moving & G, K 
& $\checkmark$ & $\sim$0.5h & 3.08\%
& $\sim$0.45h & 13.87\%
& $\sim$0.45h & 9.45\%
& \textbf{$\sim$0.03h} & \textbf{36.1\%}
\\ 
\midrule
\multicolumn{2}{c|}{
Collision Distance
}
& \multicolumn{4}{l}{
    \textbf{L}: Low collision distance at 2, 3
}
& \multicolumn{4}{l}{
    \textbf{M}: Middle collision distance at 4, 5
}
& \multicolumn{4}{l}{
    \textbf{F}: Far collision distance at 6, 7
}
\\ 
\multicolumn{2}{c|}{
Collision Speed
}
& \multicolumn{4}{l}{
    \textbf{L}: Low collision speed at 0-20
}
& \multicolumn{4}{l}{
    \textbf{M}: Middle collision speed at 20-40
}
& \multicolumn{4}{l}{
    \textbf{H}: High collision speed at 40-60
}
\\
\multicolumn{2}{c|}{
Collision Angel
}
& \multicolumn{4}{l}{
    \makecell[l]{
    \textbf{N}: Negative collision angle.
    \\
    The right corner of EV collided with the NPC.
    }
}
& \multicolumn{4}{l}{
    \makecell[l]{
    \textbf{0}: Collision angle near 0. 
    \\
    The center of EV collided with the NPC.
}
}
& \multicolumn{4}{l}{
\makecell[l]{
    \textbf{P}:Positive collision angle.
    \\
    The left corner of the EV collided with the NPC.
}
}
\\
\multicolumn{2}{c|}{
Attack Results
}
& \multicolumn{4}{l}{\textbf{G}: Slightly grazed}
& \multicolumn{4}{l}{\textbf{K}: Knock down}
& \multicolumn{4}{l}{\textbf{F}: Sent flying}
\\
\bottomrule
\end{tabular}%
}
\label{tab:discovered_ICs}
\vspace{-5mm}
\end{table*}

\section{Evaluation}

\label{sec:eval}
We implemented \sysname, a black-box fuzzer with a step-wise mutation approach starting from the collision space to find the ICSs in the AD simulator effectively.
We will release our code and prototype after the paper is accepted.
Based on the implementation, we design a series of evaluations to demonstrate the effectiveness of \sysname by investigating the following research questions:

\begin{itemize}[leftmargin=*]
\item \textbf{RQ1:} Can \sysname effectively discover ICSs?
\item \textbf{RQ2:} Can \sysname efficiently discover ICSs starting from the collision scenario space?
\item \textbf{RQ3:} Can step-wise mutation generate ICSs effectively?
\end{itemize}

To validate whether \sysname can effectively find ICSs compared to existing ADS testers (\textbf{RQ1}), we first run a state-of-the-art ADS fuzzer, DriveFuzz, equipped with our ICS oracle, and different variants of \sysname to compare their ICS identification ability with \sysname in \S~\ref{sec:eval:rq1_drivefuzz_comparison}.
To validate our intuition that starting discovery from $\mathcal{S}_{C}$ is more efficient than from the $\mathcal{S}_{NC}$ (\textbf{RQ2}), we compare the discovery time and number of ICS between DriveFuzz, \sysname starting from the collision scenario space, and \sysname in \S~\ref{sec:eval:rq2_search_space}.
Furthermore, we conduct ablations to validate whether the mutation strategy in \sysname can effectively detect ICSs (\textbf{RQ3}).
We compare the results of \sysname with \sysname using a random mutation strategy.
We also analyze the ICSs discovered by \sysname to discern if the search directions align with our designated approach in \S~\ref{sec:eval:rq3_search_direction}.
Additional ablation results, including the effect of the oracle threshold on ICS discovery, are available on our website.
\ignore{
Finally, we study the precision of our oracle in indicating ICSs (\textbf{RQ4}). We vary the threshold values of the oracle to investigate their impact on the precision and recall in \S~\ref{sec:eval:rq4_oracle-threshold}.
}

\textbf{Benchmark.} The simulator used for testing in this study is Carla, one of the most widely used autonomous driving simulators.
While various types of driving simulators exist, some are proprietary and charged, such as PreScan~\cite{prescan}.
On the other hand, other simulators primarily focus on traffic flow simulation and lack a realistic 3D environment, such as SUMO~\cite{sumo} and WayMax~\cite{waymax}.
Among the available options, LGSVL~\cite{lgsvl} and Carla~\cite{carla} provided a 3D realistic environment and multiple APIs for different ADS, such as AutoWare~\cite{autoware}, Apollo~\cite{apollo}.
However, LGSVL has suspended active development as of January 1, 2022.
Therefore, this paper focuses on testing the latest stable version of Carla (0.9.13).

\textbf{Baseline.} 
The majority of simulation-based ADS testers do not make their project code open-source.
Among the open-source projects, LGSVL was used in most of them to test Apollo before 2022, but unfortunately, they are outdated along with LGSVL.
DriveFuzz~\cite{drivefuzz} and ADFuzz~\cite{ADfuzz} use Carla as the testing platform and are open-source.
However, ADFuzz does not employ Carla's docker version, which cannot run on a server.
We have contacted the authors of ADFuzz. Unfortunately, this tool is no longer maintained. 
In contrast, DriveFuzz provided a docker version for evaluation.
Therefore, we choose DriveFuzz\cite{drivefuzz} as our baseline.
Moreover, we conducted a comprehensive comparison by evaluating \sysname starting from non-collision scenarios using our guided mutation strategy, referred to as \sysnamenc, and \sysname starting from collision scenarios using a random mutation strategy, referred to as \sysnamerandom, as our baseline.

\textbf{Configuration.} We conducted the evaluation on an Ubuntu 18.04 server powered by an 80-core Intel Xeon Gold 6242R CPU, 188 GB memory, and 2 Nvidia A30 graphics cards. 


\subsection{\textbf{RQ1:} Can \sysname Effectively Discover ICSs?}
\label{sec:eval:rq1_drivefuzz_comparison}
\begin{table}[!t]
\centering
\caption{Time and Proportion Ratios: Comparing DriveFuzz, ICSFuzz-NC, and ICSFuzz-Random to ICSFuzz. Time Ratio: Average discovery time as multiples of ICSFuzz. Proportion Ratio: ICSs found by ICSFuzz as multiples of others.}
\label{tab:time_proportion-ratio}
\renewcommand{\arraystretch}{0.8}
\resizebox{0.9\columnwidth}{!}{%
\begin{tabular}{@{}c|cc|cc|cc@{}}
\toprule
\multicolumn{1}{l|}{\multirow{2}{*}{\textbf{No.}}} & \multicolumn{2}{c|}{\textbf{DriveFuzz}} & \multicolumn{2}{c|}{\textbf{ICSFuzz-NC}} & \multicolumn{2}{c}{\textbf{ICSFuzz-Random}} \\
\multicolumn{1}{l|}{} & \begin{tabular}[c]{@{}c@{}}Time \\ Ratio\end{tabular} & \begin{tabular}[c]{@{}c@{}}Proportion \\ Ratio\end{tabular} & \begin{tabular}[c]{@{}c@{}}Time \\ Ratio\end{tabular} & \begin{tabular}[c]{@{}c@{}}Proportion \\ Ratio\end{tabular} & \begin{tabular}[c]{@{}c@{}}Time \\ Ratio\end{tabular} & \begin{tabular}[c]{@{}c@{}}Proportion \\ Ratio\end{tabular} \\ \midrule
01 & - & - & 7.00 & 17.65 & 2.29 & 18.2 \\ \midrule
02 & - & - & 36.58 & 3.02 & 23.97 & 3.56 \\ \midrule
03 & 72.07 & 22.26 & 6.71 & 2.80 & 3.03 & 3.14 \\ \midrule
04 & - & - & 17.34 & 4.15 & 4.56 & 3.99 \\ \midrule
05 & - & - & 30.48 & 15.92 & 5.30 & 7.2 \\ \midrule
06 & 77.52 & 11.13 & 20.06 & 5.46 & 13.79 & 10.91 \\ \midrule
07 & - & - & 27.81 & 2.54 & - & - \\ \midrule
08 & - & - & 35.36 & 3.21 & 8.59 & 3.87 \\ \midrule
09 & - & - & 20.72 & 2.43 & 8.94 & 3.28 \\ \midrule
10 & 17.72 & 11.72 & 16.08 & 2.60 & 15.77 & 3.82 \\ \bottomrule
\end{tabular}%
}

\end{table}

The main goal of \sysname is to discover ICSs efficiently.
Thus, we first compare \sysname with the state-of-the-art ADS fuzzer DriveFuzz, followed by a comparison with the variants \sysnamenc and \sysnamerandom.
We run each fuzzer for $36$ hours and compare the number of discovered ICSs, the respective collision types, and the time spent to discover them.
Since the default collision detector used by DriveFuzz cannot identify ICSs, we replace it with an IC-capable oracle, as discussed in \S~\ref{sec:overview}, for a fair comparison.

Table~\ref{tab:discovered_ICs} presents all ICS types found by \sysname, along with the corresponding values of control parameters and collision factors.
While DriveFuzz identified only 12 ICSs across 3 categories, \sysname discovered 473 ICSs spanning 10 types. 
Meanwhile, DriveFuzz requires 0.5 to 5 hours to discover an ICS, whereas \sysname takes 0.01 to 0.28 hours.
Although \sysnamenc and \sysnamerandom discovered almost all types of ICSs, they take longer and find fewer ICSs.
Specifically, \sysnamenc finds 148 ICSs within 0.01 to 1.21 hours per ICS, while \sysnamerandom identifies 65 ICSs within 0.03 to 2.13 hours. 

\ignore{\sout{Figure ~\ref{fig:eval:RQ2} (a) depicts the logarithmic-scaled time distribution of DriveFuzz and \sysname in discovering $12$ ICSs only detected by DriveFuzz.
Overall, the maximum discovery time \sysname uses is 3x shorter than the minimum discovery time DriveFuzz takes.
Furthermore, DriveFuzz exhibits relatively 20x larger median times than \sysname for detecting these ICSs. 
When examining the average computation time, \sysname still maintains a significant advantage.
Regardless of the type of the ICSs, DriveFuzz consistently requires 20\textasciitilde70x longer discovery time compared to \sysname as illustrated in Figure~\ref{fig:eval:drivefuzz_ave_times}.}}

As listed in Table~\ref{tab:time_proportion-ratio}, \sysname has a significant advantage in average computation time across various types of ICSs.
DriveFuzz takes 20\textasciitilde70x longer, \sysnamenc requires 2\textasciitilde15x times longer, and \sysnamerandom takes 7\textasciitilde36x longer time than \sysname.
The substantial disparity in computation time suggests that starting from the non-collision scenario space incurs a significant amount of additional exploration time, which is unproductive for discovering ICSs.

The developers have confirmed all the detected ICSs, with a partial of them fixed swiftly.
One of the ICSs has also been assigned the corresponding CVE ID due to its potential threats for real-world ADS.\footnote{The CVE number will be exposed after the paper review.}
\ignore{
Furthermore, within the discovered ICSs, we have identified another category of scenarios referred to as IC-prone scenarios, which arise due to the over-conservative behavior of our oracle (discussed in \S~\ref{sec:eval:rq4_oracle-threshold}).
IC-prone scenarios encompass situations where the distance between the vehicle and other traffic participants is extremely close but without actual collisions.
Due to its explicit threats to pedestrians and bicycles, we do not exclude them from ICSs discovered in DriveFuzz and \sysname.
The total number and proportion of IC-prone scenarios can be calculated using the precision obtained in \S~\ref{sec:eval:rq4_oracle-threshold}.
In the following comparison and analysis, we use the term ICSs to represent both ICSs and IC-prone scenarios.
\wei{maybe we can remove the IC-prone scenarios for the space limitation}}

\begin{tcolorbox}[before skip=5pt, after skip=5pt]
\textbf{Answer to RQ1:} 
\sysname outperforms state-of-the-art approaches in ICS detection, identifying 10\textasciitilde20x more issues with 20\textasciitilde70x speedup. Compared to ablation methods \sysnamenc and \sysnamerandom, it detects 3\textasciitilde8x more ICSs, 2\textasciitilde36x faster. Notably, our method uncovered high-impact vulnerabilities and a novel security issue, earning a CVE assignment.

\end{tcolorbox}

\vspace{-1.5mm}
\subsection{RQ2: Can \sysname Efficiently Discover ICS Starting From the Collision Scenario Space?}
\label{sec:eval:rq2_search_space}
\begin{table*}[t]
\centering
\caption{ICS Success Rate(SR) regarding different control parameters. SR: Ratio of ICSs to generated collision scenarios.}
\label{tab:sr-control_parameters}
\resizebox{\linewidth}{!}{%
\begin{tabular}{c|ccc|ccccc|clllllllll}
\toprule
\multirow{2}{*}{\textbf{Scenario}} 
& \multicolumn{3}{c|}{\textbf{Collision Distance}}
& \multicolumn{5}{c|}{\textbf{Collision Speed}}  
& \multicolumn{9}{c}{\textbf{Collision Angle}}                   
\\
                          & 2-3(\%)        & 4-5(\%)         & 6-7(\%)        & 0-10(\%) & 10-20(\%) & 20-30(\%) & 30-40(\%) & 40-50(\%) & -1(\%) & -0.75(\%) & -0.5(\%) & -0.25(\%) & 0(\%) & 0.25(\%)& 0.5(\%) & 0.75(\%) & 1(\%) \\ 
\midrule
FLB & 18.18 & 50.36 & 56.79 & 31.47 & 35.19 & 31.34 & 29.23 & 29.20 & 25.10 & 20.81 & 16.46 & 10.89 & 11.01 & 55.15 & 52.10 & 49.79 & 46.66  
\\
FLV & 0 & 10.33 & 34.76 & 13.59 & 21.28 & 24.06 & 22.52 & 15.32 & 10.20 & 16.71 & 21.70 & 13.23 & 16.92 & 14.82 & 27.76 & 23.62 & 27.35 
\\
PSF & 1.51 & 23.48 & 57.96 & 9.02 & 17.48 & 19.90 & 21.48 & 22.45 & - & - & - & - & 22.75 & 9.55 & 19.65 & 26.10 & 28.85 
\\
LC  & 0 & 0.14 & 15.10 & 6.59 & 8.29 & 10.14 & 9.16 & 9.86 & 14.84 & 13.21 & 9.91 & 4.24 & 0.45 & 3.78 & 11.52 & 15.84 & 14.48 
\\
PCF & 19.15 & 18.55 & 9.06 & 10.91 & 17.41 & 19.64 & 19.83 & 20.10 & 26.63 & 26.85 & 23.85 & 19.18 & 2.77 & 6.63 & 15.67 & 22.12 & 40.36 
\\
InC  & 0 & 0.14 & 15.10 & 6.59 & 8.29 & 10.14 & 9.16 & 9.86 & 14.84 & 13.21 & 9.91 & 4.24 & 0.45 & 3.78 & 11.52 & 15.84 & 14.48 
\\ 
\bottomrule
\end{tabular}%
}
\vspace{-5mm}
\end{table*}
We further compare \sysname with DriveFuzz and \sysnamenc to investigate the effectiveness of using collision scenarios to detect ICSs.
The results can be found in Table~\ref{tab:discovered_ICs} and Table~\ref{tab:time_proportion-ratio}.
\sysname shows superior performance, with proportions of 11.46\% to 36.1\% versus DriveFuzz's 1.03\% to 3.08\% in three ICS types, and 7.64\% to 40.95\% versus \sysnamenc's 0.52\% to 13.87\% across all types.
The proportion of identified ICSs remained 10\textasciitilde20x larger than that of DriveFuzz, which only discovered three types of ICSs.
Meanwhile, compared to \sysnamenc, \sysname identifies ICSs at a proportion that is 2.43 to 17.65 times higher. Although \sysnamenc can identify all ten types of ICSs, it requires more time due to the lower identification proportion.
The high identification proportion rate allows \sysname to detect all ten types of ICS more efficiently than both DriveFuzz and \sysnamenc.

The differences in proportion and type number suggest that initiating the exploration from the non-collision space is less effective in identifying a comprehensive range of ICSs. 
It increases the risk of overlooking ICSs and limits the overall efficiency of the discovery process.

\begin{tcolorbox}[before skip=5pt, after skip=5pt]
\textbf{Answer to RQ2:} The significantly higher ICS proportion and shorter detection time compared to DriveFuzz and \sysnamenc demonstrate the effectiveness of starting from the collision space rather than the non-collision space. 
\end{tcolorbox}


\vspace{-1mm}
\subsection{\textbf{RQ3:} Can Step-wise Mutation Generate ICSs Effectively?}

\label{sec:eval:rq3_search_direction}
To further study the effectiveness of our guided mutation strategy, we first compare \sysname with \sysnamerandom which uses a random mutation strategy starting from the collision scenarios.
The comparison results can be found in Table~\ref{tab:discovered_ICs} and Table~\ref{tab:time_proportion-ratio}. 
\sysname demonstrates a better performance with a minimum proportion of 11.46\% and a maximum of 36.1\%, while \sysnamerandom discovered only 1.05\% to 9.45\%.
\sysname identifies 3\textasciitilde18x more ICSs than \sysnamerandom and is 2\textasciitilde 15x faster, indicating the effectiveness of our guided mutation strategy.

Furthermore, we conduct an ablation evaluation for our step-wise mutation approach.
The scenarios tested are listed in \S~\ref{sec:emp:scenario_selection}.
The parameters involved are collision distance, speed, and angle, 
whose exact physical meanings are shown in Figure~\ref{figure:control_parameter}. 
The final correlation between ICS success rate and the control parameters are summarized in Table~\ref{tab:sr-control_parameters}. 
Detailed explanations of special cases and minor data fluctuations, along with their causes, are provided on our website.

\textbf{Collision Distance}: 
The final results indicate that the effective search direction for ICS searching generally involves increasing the collision distance, except for PCF, where it decreases.
The general trend aligns with our search strategy and the observed differences in SR among different speeds, demonstrating the effectiveness of our step size.
Contrary to intuition, which suggests more collisions occur at close distances, our findings reveal more ICSs when vehicles are farther apart with one changing its driving behavior. 
The trend suggests that interactions at greater distances may create more critical situations than previously assumed.
\ignore{
Overall, the data trend in the final result indicates that the effective search direction for ICS searching of collision distance is towards increasing distance in all scenarios except for PCF, where the direction decreases distance.  

In scenario FLB, FLV, LC and PSF, 
there is a significant positive correlation between ICS SR and collision distance.
Similarly, in scenario InC, the overall SR trend remains, with a slight fluctuation in the middle distance range.
In scenario PCF, 
the overall ICS SR exhibits a declining trend as the distance increases.
The reason for the negative correlation in scenario PCF is that the NPC pedestrian is walking in the direction perpendicular to the driving direction of the EV.
As the collision distance increases, the probability of the NPC pedestrian moving out of the collision area also increases. 

The overall upward trend aligns with our search direction, and the significant difference in SR observed during the transition from low speed to medium/high speed demonstrates the effectiveness of our search step size.
Moreover, the data trend contradicts the intuition regarding the relationship between collision distances and real collisions. 
Intuitively, one would expect a higher occurrence of collisions when two vehicles are close to each other and one of them changes the driving behavior, suggesting that more ICSs should occur within a close distance.
However, the results reveal that more ICSs are found when the distance between two vehicles is far, with one of them altering their driving behavior.
}

\textbf{Collision Speed}:
The experimental results indicate that although collision speed has a less significant impact compared to the other parameters, a positive correlation between ICS success rates and collision speed is evident in most scenarios. Table~\ref{tab:sr-control_parameters} shows a positive trend between collision speed and final ICS SR in most cases, with more ICSs found at middle-range collision speeds, except for FLV.

The data highlights that higher collision speeds align with our search direction and show that our steps effectively detect ICSs. Contrary to expectations, our study reveals that more collisions typically occur at low speeds. However, our results show more ICSs at middle and high speeds.
The discrepancy may be because high collision speeds reduce the time for the collision detector to react, leading to more ignored collisions.
\ignore{
The experimental results indicate that, although the collision speed parameter does not significantly impact the overall outcomes compared to the other two parameters, a positive correlation still exists between the ICS success rates and collision speed in most scenarios.
From Table~\ref{tab:sr-control_parameters}, it can be observed that in most scenarios, there exists a positive trend between collision speed and the final ICS SR.
Except for FLV, more ICS were discovered while collision speed was at the middle range.

The data illustrates that more ICSs are found when the collision speed is high, which aligns with our search direction and illustrates that our search steps effectively detect ICSs.
Our empirical study reveals that, in reality, a higher number of collisions occur at low collision speeds, where one would expect a more significant presence of ICSs. 
However, contrary to this expectation, our experimental results 
are in the opposite direction in most scenarios about the collision factor speed, which demonstrates that more ICSs are found when the collision speed is middle and high. 
The distinction may be because, at high collision speeds, the collision detector lacks sufficient time to react and detect the collision accurately, leading to a higher proportion of ignored collisions.
}


\textbf{Collision Angle}: 
ICS SR trends related to angle show high values on both ends and low in the middle. Specific environmental settings cause data fluctuations. For collision angles not equal to zero, most scenarios support our study's conclusions. Scenario PSF lacks ICS for negative angles because the pedestrian is in front of the EV. 

The search direction for collision angles is from -1 or 1 to 0 ($-90^\circ$ or $90^\circ$ to $0^\circ$ in Figure~\ref{figure:control_parameter}). 
The steep trend confirms the effectiveness of our search direction and steps. However, real collision frequencies are higher near 0 angles, where vehicles often move straight. Near-zero angles result in stronger collisions, making them less likely to be missed. Angles near ±1 leading to more identified ICSs.

\ignore{
The data trends of ICS SR related to angle in different scenarios align with the trend of high on both sides and low in the middle. 
The fluctuations in the data result from some specific environmental settings.
For collision angle values greater or smaller than 0, the ICS SR trends of most scenarios match our conclusion derived from the study. For scenario PSF, it does not have ICS when the collision angle is negative since the pedestrian is standing in the right front of the EV.

Overall, for collision angle values greater than 0, a positive correlation between the collision angle and the final ISC SR is observed in almost all scenarios except for FLV scenarios, indicating its effectiveness in ICS detection.
The data trend of scenarios in FLV with the positive angle range exhibits fluctuations. 
One reason is that many ICS-prone scenarios mentioned in \S~\ref{sec:eval:rq1_drivefuzz_comparison} happen when the angle is near 0.

For collision angle values smaller than 0, the ICS SR trends of most scenarios also match our conclusion derived from the study, except for PSF and FLV.
For scenario PSF, it does not have ICS when the collision angle is negative since the pedestrian is standing in the right front of the EV.
In scenario FLV, the \textcolor{red}{local maximum} \sout{peak} observed in the negative angle range can also be attributed to a higher number of occurrences of type 3 ICS within the specific angle range.
The decreasing trend observed as the collision angle approaches $-1$ is due to the scenarios being at the border between collision and non-collision, leading to a decrease in both the number of collisions and the corresponding number of ICSs.

The search direction for collision angle is from -1 or 1 to 0 ($-90^\circ$ or $90^\circ$ to $0^\circ$ in Figure~\ref{figure:control_parameter}).  
The steep slope of the data trend demonstrates the effectiveness of the search direction and search steps obtained from our study.
However, the collision frequency trend in reality is the opposite; there are more collisions when the collision angle is close to 0, given that the "Vehicle State While Colliding" has the highest proportion of vehicles moving straight ahead.
The reason for such counter-intuition distinction is that collision angles near 0 often result in a stronger collision force, making them less likely to be missed by the collision detector. 
Conversely, collision angles approaching 1 and -1 ($-90^\circ$ and $90^\circ$) encompass a broader range of collision and non-collision scenarios, leading to more identified ICSs.
}


\begin{tcolorbox}[
float=true,
float=htbp,
before skip=5pt, after skip=5pt
]
\noindent \textbf{Answer to RQ3:}
The comparison results with \sysnamerandom validate our guided mutation strategy.
Data trends for all three control parameters and ICS SR support the effectiveness of the step-wise method conclusions of our empirical study in \S~\ref{sec:emp_study}.
Together with our selected scenarios, 
we avoid generating numerous irrelevant scenarios to detect ICSs efficiently.
Moreover, the ICS trend in the simulator is counter-intuitive to real collisions, highlighting the challenge of detecting ICSs with existing efforts and the potential risks that cannot be overlooked.
\ignore{
The comparison results with \sysnamerandom indicate the effectiveness of our guided mutation strategy. 
The data trends of all three control parameters with their ICS success rate underscore the effectiveness of the step-wise method and demonstrate the conclusions of our empirical study in \S~\ref{sec:emp_study}.
Together with our selected scenarios, we avoid generating numerous irrelevant scenarios to detect ICSs efficiently.
Meanwhile, these general factors also help us to analyze the reason for ICS occurrences. 
Moreover, the trend of discovered ICSs in the simulator is counter-intuitive to real collisions.
Such discrepancy highlights the challenge of detecting ICSs with existing efforts and the potential risks that cannot be overlooked.}

\end{tcolorbox}

\ignore{
\subsection{\textbf{RQ4:} Can \sysname Precisely Identify \textcolor{red}{\sout{All Detected}} ICSs?}
\label{sec:eval:rq4_oracle-threshold}


\begin{figure}[t]
  \begin{subfigure}[t]{0.19\textwidth}
    \includegraphics[width=\textwidth]{fig/result/oracle_threshold/threshold_precision_flb.png}
    \caption{Follow Leading Bicycle - Precision and Recall}
    \label{fig:oracle_threshold-precision-flb}
  \end{subfigure}
  \hfill
  \begin{subfigure}[t]{0.19\textwidth}
    \includegraphics[width=\textwidth]{fig/result/oracle_threshold/threshold_num_flb.png}
    \caption{Follow Leading Bicycle - Number of TP}
    \label{fig:oracle_threshold-num-flb}
  \end{subfigure}

  \begin{subfigure}[t]{0.19\textwidth}
    \includegraphics[width=\textwidth]{fig/result/oracle_threshold/threshold_precision_flv-h.png}
    \caption{Lane Change - Precision and Recall}
    \label{fig:oracle_threshold-precision-flv-h}
  \end{subfigure}
  \hfill
  \begin{subfigure}[t]{0.19\textwidth}
    \includegraphics[width=\textwidth]{fig/result/oracle_threshold/threshold_num_flv-h.png}
    \caption{Lang Change - Number of TP}
    \label{fig:oracle_threshold-num-flv-h}
  \end{subfigure}
  
  \begin{subfigure}[t]{0.19\textwidth}
    \includegraphics[width=\textwidth]{fig/result/oracle_threshold/threshold_precision_pcf.png}
    \caption{Pedestrian Standing Front - Precision and Recall}
    \label{fig:oracle_threshold-precision-pcf}
  \end{subfigure}
  \hfill
  \begin{subfigure}[t]{0.19\textwidth}
    \includegraphics[width=\textwidth]{fig/result/oracle_threshold/threshold_num_pcf.png}
    \caption{Pedestrian Standing Front - Number of TP}
    \label{fig:oracle_threshold-num-pcf}
  \end{subfigure}
  \caption{\S~\ref{sec:eval:rq4_oracle-threshold}: The effect of oracle threshold values.}
  \label{fig:oracle_threshold}
\end{figure}

As we use the object detection bounding box as our oracle for ICS discovery, we aim to investigate whether this oracle is sufficiently reliable and accurate in identifying ICSs.
Therefore, we conducted an ablation study to examine the relationship between threshold and performance and investigate the impact of the IoU threshold on the final ICS results.
Three representative scenarios, FLB, LC, and PSF, are tested, each involving different struck objects.
We explore whether different NPC types affect the TP proportion when the IoU threshold changes.
 
For scenario FLB, 
We set the collision distance to 3 and the collision speed to 2. The collision angle on the x-axis was fixed at 0.02, while the collision angle on the y-axis varied from 0 to 1 in increments of 0.02. 
Similarly, for scenario LC, we set the collision distance to 5 and the collision speed to 30. The collision angle on the y-axis was fixed at 0.02, and the collision angle on the y-axis ranged from 0 to 1 with a step of 0.02. 
For scenario PSF, we set the collision distance to 2 and the collision speed to 4. The collision angle on the x-axis was fixed at 0, and the collision angle on the y-axis ranged from 0 to 1 with a step of 0.02. 
Each scenario will be run 50 times under the specification above.
The final results have been manually confirmed and presented in Figure~\ref{fig:oracle_threshold}.

The final precision, recall, and the number of true positives are shown in Figure~\ref{fig:oracle_threshold}.
For scenario FLB, involving a bicycle as the NPC, the precision remains relatively consistent for thresholds ranging from 0 to 0.15. However, the recall decreases as the threshold increases, with a significant drop beyond 0.15. The optimal threshold value for FLB is still 0. 
Similarly, for scenario LC with a vehicle as the NPC, the precision is high at thresholds of 0 and 0.2. However,  there is only 1 TP when the threshold is 0.2, the number of true positives exhibits a decreasing trend, and an IoU threshold of 0 yields the best results. In scenario PSF with a pedestrian as the NPC, raising the threshold above 0.05 did not result in additional ICSs. Therefore, setting the threshold at 0 is still the most reasonable choice for our ICS discovery method.

\begin{tcolorbox}
\textbf{Answer to RQ4:} Despite the potential false positives in the threshold set for our oracle, \sysname remains highly effective in detecting ICSs according to our empirical evidence.
Additionally, the false positive results may include IC-prone scenarios that hold practical significance, introducing new threats for ADS in real-world scenarios.
\end{tcolorbox}
}

\vspace{-3mm}
\section{Related Work}
\label{sec:related_work}




\textbf{Simulation-based ADS testing.}
The popular adopted ADS testing methodology involves simulation-based testing, real-world vehicle testing, and hardware-in-the-loop testing.
Simulation-based ADS testing~\cite{zhou_specification,tang_systematic,avfuzzer,mosat,song2023discovering,wachi-ads-rl,planfuzz,drivefuzz,avchecker} is widely adopted due to the advantage of high-fidelity, cheap and fast.
The specific works differ primarily in their deployment methods, including the specific mutation strategy they employ, such as a genetic algorithm~\cite{mosat,avfuzzer} and reinforcement learning~\cite{wachi-ads-rl}, and the criteria they use to estimate the danger level of the generated scenarios, such as the distance between traffic participants and the number of hard brakes~\cite{planfuzz,avfuzzer,drivefuzz,avchecker}.
However, all of these works focus on improving the speed and effectiveness of ADS testing, overlooking the reliability of their testing platform, which is the ICSs in this paper. 
Therefore, our work can provide a more solid platform for future ADS testing for more reliable results.

\textbf{Traditional fuzzers.}
Fuzz testing has recently witnessed its prosperity in detecting security flaws by generating enormous test cases and monitoring the executions for bug detections. 
The current dominant approach of fuzzing for test case generation is mutation-based strategy, which introduces minor changes to existing interested inputs to keep the input valid yet exercise new behavior.
Various fuzzers~\cite{AFLplusplus-Woot20,alfgo,afl} leverage mutation-based input generation strategy to detect generic software bugs leading to software crashes or vulnerabilities.
Additionally, traditional fuzzers mutate interested inputs following the guidance of code coverage~\cite{afl,AFLplusplus-Woot20} or predefined distance metrics~\cite{alfgo}.
However, they cannot directly apply to test ICS since ICS is a kind of semantic error rather than the logical error that traditional fuzzers can handle using a simple oracle (e.g., software crash).
Nevertheless, for ICS testing, it is still possible to inherit the mature generation insight for better detecting ICS with better integration in future work.

\ignore{
Another two commonly used methods are formal verification and attack-based testing. 
Formal verification focused on proving the safety of ADS in its operation design domain, 
primarily examining the logic and code implementation of the system itself. 
AV-Checker\cite{avchecker} combines the static analysis and Z3 SMT solver. 
AV-Checker generates counterexamples to detect inconsistencies between system performance and requirements by comparing these specifications with public rule specifications. 
Attack-based testing aims to create adversarial examples that deceive the sensing system of an attacked vehicle or change the internal state of the attacked vehicle. 
The ultimate goal is to get the vehicle to engage in dangerous driving behaviors such as hard braking or sharp turns.

In \cite{nonadaptive-testing}, human-reported crash scenarios from NHSTA collision reports are automatically converted into test scenarios to evaluate the ADS to see if it performs well under accident-prone environmental conditions. 
However, it does not utilize the test feedback to generate new test cases to test the system and may ignore the difference between a human driver and an ADS. 
Common simulator-based scenario testing employs one-dimensional or multidimensional fitness functions to score each testing round and generate new test cases based on these fitness scores. These methods typically have a task-specific oracle. 
}

\ignore{
\noindent \textbf{Consistency Checking }
\label{sec:related_work_fuzzingADS2}
Consistency checking is a widely studied subarea in software engineering. 
In 1994, the concept of "viewpoint" was introduced by \cite{1994inconsistency}, which combines various aspects such as 
work plans, work domains, and specifications from different stakeholders during the software development process. 
The inconsistency checking here becomes a logical comparison of the viewpoint.
Several research papers explored this area in the next years using different approaches. \cite{2001inconsistency_survey} divided them into four categories: model checking, specialized algorithm, logical inference, and theorem proving.

Model checking involves utilizing or combining existing model checkers to perform consistency checking. 
Specialized algorithms are designed specifically for analyzing models and detecting inconsistencies. Logical inference employs formal inference techniques to derive inconsistencies, while theorem proving utilizes theorem provers for reasoning about inconsistencies.
In \cite{consistency_checking-review}, a proper definition of consistency types of software models is proposed, including exogenous horizontal and exogenous vertical consistency. Exogenous horizontal consistency refers to the consistencies among different models or models against rules and constraints manually specified from other artifacts. Vertical consistency, in contrast, includes the consistencies among the same type of models or the consistency between a design model and a corresponding implementation.

In recent years, with the development of drones and autonomous driving, consistency-checking work has appeared in these domains. CP-inconsistency\cite{cp-incon} identified numerous cyber-physical inconsistency vulnerability cases in UAV control systems.
Then, researchers constructed a virtual test platform where virtual vehicles operated in an ideal world without obstacles. 
Additionally, they modeled the UAV control system through System Identification and divided the control model into different stages like taking off, hanging, etc. 
By comparing the control commands in both the ideal world and real-world scenarios and incorporating the distance in the control flow graph as a multi-objective optimization problem to solve, they identified Pareto points as inconsistency vulnerabilities.
However, the solution cannot be directly applied to autonomous driving since conditions are way more complex.  

}

\vspace{-1mm}
\section{Conclusion}

In this paper, we are the first to systematically study an overlooked reliability problem in modern autonomous driving simulators. 
Furthermore, we propose a black-box fuzzing approach to efficiently search ICSs by initiating the search process from empirically studied representative collision scenarios with a stepwise mutation approach.
The evaluation results demonstrate that we can effectively discover ICSs significantly outperform the state-of-the-art ADS tester, DriveFuzz.
Our tool, \sysname, finds 10\textasciitilde20x more ICSs with a 20\textasciitilde70x speedup and uncovers seven additional types of ICSs that DriveFuzz fails to find.
Moreover, the developers confirmed all the ICS issues as bugs; one of them is too severe, which can cause a real-world impact, and thus, is assigned with CVE ID.
Our findings and insights can shed light on the future RE security of autonomous driving simulators.



\bibliographystyle{IEEEtran}
\bibliography{refs}



\end{document}